\documentclass[12pt]{article}
\usepackage{epsfig}
\usepackage{amsfonts}
\usepackage[tbtags]{amsmath} 
\usepackage{amsthm} 
\usepackage{times}
\usepackage{amsmath}%
\usepackage{amssymb}%
\usepackage{graphicx}
\usepackage{epstopdf}
\usepackage{rotating}
\usepackage{booktabs}
\usepackage{caption}
\usepackage[mathscr]{euscript}
\usepackage{lscape}
\usepackage{subfigure}
\usepackage{subcaption}
\usepackage{mathrsfs}
\usepackage{subcaption}
\usepackage{times}
\usepackage{rotating}
\usepackage{float}
\usepackage{afterpage}
\usepackage{url}
\usepackage{xcolor,colortbl}
\usepackage{geometry}
\usepackage{multirow}
\usepackage{graphics}
\usepackage{dsfont}
\usepackage{bm}
\usepackage{enumitem}
\usepackage{url}

\usepackage{float}
\usepackage{authblk}
\usepackage{adjustbox}

\definecolor{Gray}{gray}{0.85}
\definecolor{LightCyan}{rgb}{0.88,1,1}
\usepackage{natbib}
\usepackage{setspace,xr}
\usepackage[implicit=false]{hyperref}

\newtheorem{thm}{THEOREM}[section]

\newtheorem{lemma}{LEMMA}[section]

\newtheorem{remark}{REMARK}[section]
\newtheorem{definition}{DEFINITION}[section]
\renewcommand{\theequation}{\thesection.\arabic{equation}}
\newtheorem{corol}{Corollary}[section]
\newtheorem{propos}{Proposition}[section]
\newtheorem{assumption}{ASSUMPTION}[section]

\def\captionof#1#2{{\def\@captype{#1}#2}}
\makeatletter
\renewcommand\paragraph{\@startsection{paragraph}{4}{\z@}%
	{-3.25ex\@plus -1ex \@minus -.2ex}%
	{1.5ex \@plus .2ex}%
	{\normalfont\normalsize\bfseries}}
\makeatother
\renewcommand\thetable{\thesection.\arabic{table}}

\usepackage{tikz}
\usepackage{quantikz}
\usetikzlibrary{trees}
\usetikzlibrary{positioning}
\usetikzlibrary{arrows.meta}
\usetikzlibrary{positioning}
\usetikzlibrary{decorations.markings}
\usetikzlibrary{decorations.text}
\usetikzlibrary{matrix}
\usepackage{float}
\usepackage{authblk}
\usepackage{adjustbox}

\def\hilbert{\mathbb{H}}
\def\complex{\mathbb{C}}

\begin{document}
	\title{On Quantum and Quantum-Inspired Maximum Likelihood Estimation and Filtering \\ of Stochastic Volatility Models\footnote{The first author acknowledges the financial support from an IBM Global University Program Academic Award and an IonQ QLAB Global Users Program grant. We thank Andrii Babii, Peter Hansen, Vanio Markov, Valentin Verdier and Stefan Woerner and seminar participants at Bocconi University and UNC Chapel Hill for comments on our paper. We also thank participants at the Macro and Financial Econometrics Workshop, Heidelberg University, the Quantum Computing Applications in Economics and Finance Workshop at the University of Pennsylvania and the SAS Quantum Workshop for their feedback. }}
\author[1]{Eric Ghysels}
\author[1]{Jack Morgan}
\author[2]{Hamed Mohammadbagherpoor} 
\affil[1]{University of North Carolina, Chapel Hill NC, USA}
\affil[2]{IBM T.J.\ Watson Research Center, Yorktown Heights, New York, USA}
\date{First draft: August 1, 2024\\
	This version: \today}
\maketitle

\begin{abstract}
	\noindent Stochastic volatility models are the backbone of financial engineering. We study both continuous time diffusions as well as discrete time models. We propose two novel approaches to estimating stochastic volatility diffusions, one using Quantum-Inspired Classical Hidden Markov Models (HMM) and the other using Quantum Hidden Markov Models. In both cases we have approximate likelihood functions and filtering algorithms that are easy to compute. We show that the non-asymptotic bounds for the quantum HMM are tighter compared to those with classical model estimates.  
\end{abstract}	
\thispagestyle{empty}
\setcounter{page}{0}	
\newpage

\newpage
\section{Introduction}

Continuous time diffusion models with stochastic volatility are the backbone of a vast literature on continuous time asset pricing models, see e.g.\ \cite{mertoncontinuous}. Stochastic volatility (SV) accounts for the changing uncertainty or risk over time, making the models more robust in predicting asset behavior. Such models are particularly valuable in option pricing, risk management, and portfolio optimization. 

\medskip

Maximum Likelihood Estimation (MLE) of parameters with discrete time sampled observations is confined to models without latent volatility. Even then, a closed form solution for the likelihood function is only available in a handful of cases where one can solve a so called Fokker-Planck partial differential equation to obtain the transition density - see \cite{lo1988maximum} for details. In all other cases one has to resort to numerical methods of which there are many, see e.g.\ \cite{ait2002maximum,ait2002telling}, \cite{durham2002numerical}, \cite{pedersen1995consistency}, among others.

\medskip

In the more realistic setting of time-varying latent volatility there is the additional complication that besides parameter estimation, ideally we also want to filter the latent volatility process. 
Several sophisticated econometric techniques have been proposed with Bayesian inference via Markov Chain Monte Carlo (MCMC) simulations the most commonly used approach.  The empirical implementation of such models has been studied extensively and still poses many challenges, see e.g.\ \cite{ghysels19965} for an early review of the literature and \cite{dinunno2023constant} for a recent survey.\footnote{While our main focus will be continuous time diffusion models, the case of discrete time SV models is also covered by our analysis. Since continuous time diffusion models are more challenging, they get most of our attention.}

\medskip

We propose two novel approaches to estimating stochastic volatility diffusions, one using Quantum-Inspired Classical Hidden Markov Models (HMM) and the other using Quantum Hidden Markov Models. In both cases we have approximate likelihood functions and filtering algorithms that are easy to compute.  The idea to explore connections between continuous time SV diffusions and discrete time HMM has been pursued by \cite{genon2000stochastic} who construct a hidden Markov model for discretely observed returns with a bivariate latent Markov process representation of integrated and spot volatility generated by a continuous time SV diffusion. However, the Markov chain does not have a closed form solution and therefore \cite{genon2000stochastic} resort to moment-based estimators. As a result, their estimator is potentially inefficient and there is no filtering algorithm for the hidden volatilities. 

\medskip

Our paper offers several contributions. First, we formulate what we call a quantum-inspired classical HMM approximation to discretely sampled SV diffusion models. The approximation comes from the fact that the latent volatility process is defined over a finite set of hidden states instead of the continuous valued hidden process.\footnote{\cite{rossi2006volatility}, among others, formulate discrete time SV models as hidden Markov models with potentially a large number of hidden states. Their model is not formulated as a discretization of a SV diffusion model, but instead is stated in terms of a discrete time data generating process. Our paper also builds on earlier work by \cite{ghysels2023quantum} who proposed quantum computational algorithms for a class of regime switching volatility model driven by a Markov chain with observable states.}  By quantum-inspired we mean that the design is built on a computational setup amenable to quantum hardware. The classical HMM yields parameter estimates as well as a filtering algorithm for volatility. In addition, it yields an approximate likelihood function. The approximation error is reduced by increasing the number of discrete hidden states. This brings us to the second contribution pertaining to quantum HMM. In particular, if $n_L^c$ is the number of discrete hidden volatility states in a classical HMM, we achieve the same accuracy in terms of Kullback-Leibler divergence, with $\sqrt{n_L^c}$ quantum states.  Hence, there is a quadratic speedup in the number of states it takes to achieve the same approximation to the data generating process. It is one of several examples where quantum computing, which opens new research opportunities for econometricians, has advantages over classical counterparts. For a given finite number of Markov chain states, the quantum informational notions of superposition and entanglement provide a richer class of Quantum Hidden Markov models compared to classical HMM, see \cite{accardi2023hidden} for a general discussion.

\medskip

Third, \cite{lehericy2021nonasymptotic} derives non-asymptotic bounds for approximate likelihood estimators for HMM. Using these bounds, we show that the quantum HMM yields tighter non-asymptotic bounds on the parameter estimates and filtered volatilities compared to its classical counterpart. Fourth,  the operator-theoretic foundations of quantum computing/mechanics provides elegant formulas for filtering and estimation using tools, such as partial trace functions and Kraus operators, typically not used by econometricians.  Finally, we implement the proposed methods with real data. The implementation of the proposed methods is still challenging with the current state of hardware, but rapid technological progress is bringing these methods closer to reality, and the algorithms we introduce readily expand to technological innovations in quantum computing.

\medskip

It is also worth noting that the paper is about {\it classical} statistical inference,  and not about a comparison with  {\it Bayesian} algorithms based on Markov Chain Monte Carlo simulations involving  Metropolis-Hastings accept/reject draws, see e.g.\ \cite{jacquier1994bayesian}. It would be difficult to engage in a comparative analysis of the two fundamentally different procedures. The quantum(-inspired) algorithms provide quasi-closed form solutions to filtering and the likelihood function to estimate parameters. It does not involve the specification of priors, nor the use a a candidate density required for the accept/reject routine. It is therefore not the purpose of our paper to make a direct comparison with Bayesian MCMC as this would not be straightforward. Instead, our paper focuses on comparing classical statistical methods for estimating stochastic volatility diffusions and exploring quantum computational advantages. 
\medskip

Our paper contributes to a growing literature pertaining to quantum computing (QC) advances. Financial applications include: (a) quantum Monte Carlo simulation where a potential of quadratic speed-up has been pursued, see e.g.\ \cite{woerner2019quantum}, (b) exponential speed-ups in solving dynamic asset pricing models, see e.g.\ \cite{ghysels2024quantum}, (c) dynamic programming applied with economic and finance applications, see e.g.\ \cite{fernandez2022dynamic}, (d) option pricing, see e.g.\  \cite{stamatopoulos2020option} for Black-Scholes type models with exotic option pricing. See \cite{kaneko2022quantum}, \cite{vazquez2021efficient}, \cite{ghysels2023quantum}, for option pricing models with stochastic volatility, among others, among others.\footnote{There are also a number of recent literature surveys, including \cite{orus2019quantum} and \cite{herman2022survey}, in addition to the textbook by \cite{hull2020quantum}.
}

\medskip

The paper is structured as follows. Section \ref{sec:qinspired} presents the quantum-inspired classical hidden Markov model approach to stochastic volatility diffusion models. Section \ref{sec:QHMM} presents the quantum HMM counterpart followed by Section \ref{sec:dimapprox} where we elaborate on approximation and quantum dimensionality reduction.
Section \ref{sec:implementation} contains the empirical implementation and Section \ref{sec:nonasymp} compares the classical and quantum HMM in terms of non-asymptotic bounds.
Conclusions follow. Appendices cover regularity conditions, technical lemmas, proofs for theoretical results in the paper, details about the empirical implementation and finally a summary of some useful concepts of quantum information theory.

\section{A Quantum-Inspired Classical Approach \label{sec:qinspired}}

We start with a univariate stochastic volatilty diffusion for $\{Y_\tau: \tau \in \mathcal{T} \subseteq \mathbb{R}^+, y_\tau \in S_Y \subseteq \mathbb{R}\}$ characterized by the following stochastic differential equation:
\begin{eqnarray}
	\label{eq:SDE}
	d Y_\tau & = & \sigma_\tau d B_\tau \\
	 d V_\tau & = & a(V_\tau) d\tau + b(V_\tau) d W_\tau \qquad \text{with } V_\tau = \sigma_\tau^2 \notag
\end{eqnarray}
with $(B_\tau, W_\tau)_{\tau \geq 0}$ two independent standard Brownian motions, $V_\tau$ a positive process and $V_0$ independent of $(B_\tau, W_\tau)_{\tau \geq 0}.$  The drift $a(\cdot)$ and diffusion $b(\cdot)$ functions satisfy standard regularity conditions appearing as Assumptions \ref{assum:A0} through \ref{assum:a3} in Appendix Section \ref{appsec:regcond}.

\medskip

We don't observe the continuous path of the process $Y_\tau$ and only have a record of discrete time $y_t$ $\equiv$ $Y_{t \Delta}$ for some fixed time increments $\Delta,$ yielding $\{y_t: t \in T \subseteq \mathbb{N}, y_t \in S_y \subset \mathbb{R}\}$.\footnote{With some additional notation we can accommodate unequally spaced discrete time observations, but for simplicity we use a fixed sampling scheme.} The process $y_t$ is non-Markovian due to the latent volatility process. However, as Theorem 3.1 of \cite{genon2000stochastic} - henceforth GJL - shows, under regularity conditions \ref{assum:A0} through \ref{assum:a3}, the following is a hidden Markov model with continuous state space:
\begin{eqnarray}
	\label{eq:MarkovSV}
	\Delta y_t & := & \int_{(t-1)\Delta}^{t\Delta} \sigma_\tau d \tau \\
	\mathfrak{U}_t & := & (\bar{V}_t, V_{t \Delta}) \qquad \bar{V}_t := \int_{(t-1)\Delta}^{t\Delta} V_\tau d \tau. \notag 
\end{eqnarray}
The process appearing in equation (\ref{eq:MarkovSV}) is a hidden Markov model (HMM), with integrated $\bar{V}_t$ and spot $V_{t \Delta}$ volatilities as discrete time continuous valued joint latent processes and (log) returns $\Delta y_t$ as observable states. Note that $\Delta y_t$ $\sim$ $\mathcal{N}(0,\bar{V}_t )$ conditional on $\mathfrak{U}_t.$ However, $\bar{V}_t$ is non-Markovian, while $V_{t\Delta}$ is Markovian, and therefore we need the joint process to build the HMM. Moreover, the transition kernel for $\mathfrak{U}_t$ is unknown, which is why GJL opted to estimate parameters embedded in the drift $a(\cdot)$ and diffusion $b(\cdot)$ functions using moment-based estimators.
To proceed with the hidden states we need Lemma	\ref{lemma:GJL} -  which is part of the proof of Theorem 3.1 of GJL, namely (for completeness we  provide a proof in Appendix Section \ref{appsec:lemmas}) - which states that
under Assumptions \ref{assum:A0} through \ref{assum:a3}, the hidden discrete time Markov process $\mathfrak{U}_t$  := $(\bar{V}_t, V_{t \Delta})$ defined in equation (\ref{eq:MarkovSV}) has a time homogeneous transition kernel dependent only on $V_{t \Delta}.$ Moreover, the process $\Delta y_t$ is strictly stationary and ergodic. 

\medskip

\noindent Henceforth, we will adopt the terminology of the HMM literature, notably calling $\Delta y_t$ $\sim$ $\mathcal{N}(0,\bar{V}_t )$ the emission density conditional on the hidden state, namely:

\begin{definition}[Data generating process]
	\label{def:targetHMM} The process $\{\Delta y_t, \mathfrak{U}_t \}$ is a time homogeneous Markov process defined on $(\Omega, \mathcal{F}^{\Delta y}_{t \in T}, \mathcal{F}, \mathbb{P}^*_{\Delta y, U})$ with hidden Markov chain state space $(U,\mathscr{B}(U))$ and transition kernel $\mathscr{A}^*(U,U^\prime).$ Observations take values in Polish space $S_{\Delta y},$ endowed with a Borel probability measure $\mathbb{P}^*$ [henceforth all starred symbols pertain to population entities]. For each hidden state $U$ there is an emission density $\mathcal{P}^*(\cdot \vert U).$ 
\end{definition}

In this paper we consider two types of estimation approaches, one we will call parametric and the other non-parametric. The former involves parametric specifications for $a(\cdot)$ and  $b(\cdot).$ The focus will be on models with closed form transition densities for spot volatility. The goal will be to formulate a MLE not only of $a(\cdot)$ and  $b(\cdot)$ but also filtering of the latent integrated $\bar{V}_t$ and spot $V_{t \Delta}$ volatilities. For the non-parametric approach we approximate the transition kernel $\mathscr{A}^*(U,U^\prime)$ instead. This will also yield filtered estimates of integrated and spot volatility in addition to an estimate of the transition kernel approximation.

\begin{remark}
	\label{remark:discrSV}
	So far we focused on continuous time diffusion models. Our analysis also applies to the large class of discrete time models (see \cite{ghysels19965} for many examples). Succinctly this class of models can be written as:
	\begin{eqnarray}
		\label{eq:discreteSV}
		\Delta y_t & = & \sigma_{t} \varepsilon_t \qquad \varepsilon_t \text{ i.i.d. } \mathcal{D}(0,1) \\
		\sigma_{t}^2 & = & \text{Markov}_+(\sigma_{t-1}^2,\text{parameters}) \notag \\
		\text{or } 	\log \sigma_{t} & = & \text{Markov}(\log \sigma_{t-1},\text{parameters}) \notag
	\end{eqnarray}
	where $\mathcal{D}(0,1)$ is a distribution with mean zero and variance one, possibly but not necessarily Gaussian. Markov$_+$ is a positive valued Markov process.
\end{remark}
\subsection{Classical HMM Formulation}

For practical reasons, with the ultimately focus on computational issues, we will restrict our attention to compact sets for building models, although the theory goes through without this restriction. More specifically, for all $t:$ (a) $\Delta y_t$ $\in$ $S_{\Delta y},$ a compact subset of $\mathbb{R},$ (b) $V_t$ $\in$ $S_V$ and $\bar{V}_t$ $\in$ $S_{\bar{V}},$ both compact subsets of $\mathbb{R}_+.$  
Since our ultimate goal is to formulate quantum computational approximate likelihood functions we also need to modify the observable states. In this section we start with a classical HMM formulation before introducing the quantum HMM. The classical HMM is quantum inspired, which is novel to the literature. To that end, to unburden the notation we set $\Delta$ = 1 and suppose that the
process $\Delta y_{t}$ is recorded across discrete bins, more specifically:
\begin{definition}[Obsersable states]
	\label{def_obs}
	Let $S_{\Delta y}$ be a compact set with partition  $\mathcal{B}_O$ := $\{b_1, \ldots, b_{n_O}\}$ such that $\bigcup_i b_i$ = $S_{\Delta y}$ and $b_i$ $\cap$ $b_j$ = $\emptyset$ for $i$ $\neq$ $j$ with $n_O$ the number of bins, i.e.\ $n_O$ := $\vert \mathcal{B}_O \vert.$ The maximal length of the bins is denoted by $\ell_{max}^b.$ Then, we have a discrete time and state process $\{\mathfrak{y}_{t}: t \in T \subset \mathbb{N}, \Delta y_t \in b_i \text{ then } \, \mathfrak{y}_{t} = \iota_i \}$ where $\iota_i$ is the $i^{th}$ coordinate vector of $\mathbb{R}^{n_o}.$ 
	We define the set $\mathcal{S}_O$ := $\{\iota_i, i = 1, \ldots, n_O\}.$	
	For any given $\mathfrak{U}_t$ we define the emission densities as:
	\begin{equation}
		\label{eq:dirac}
		\mathcal{P}_{\mathfrak{y}}(\mathfrak{y}_{t+1} := \iota_i \vert \mathfrak{U}_t) = \int_{S_{\Delta y}} \mathcal{P}(x \vert \mathfrak{U}_t) \delta_{b_i}(x) d x \qquad \forall \iota_i \in \mathcal{S}_O,
	\end{equation}
	with $\delta_{b_i}$ the Dirac delta function associated with bin $b_i$ and $\mathcal{P}(x\vert \mathfrak{U}_t)$ the Gaussian probability distribution associated with $\mathcal{N}(0,\bar{V}_t )$ and $\bar{V}_t$ driven by $V_t.$
\end{definition}

\begin{remark}
	\label{remark:emissiondensity}
	The setup appearing in equation (\ref{eq:MarkovSV}) features emission densities that are known, i.e.\  $\Delta y_t$ $\sim$ $\mathcal{N}(0,\bar{V}_t )$ conditional on $\mathfrak{U}_t.$ Therefore, there is no distinction between the population emission density $\mathcal{P}^*$ versus a potential approximate density $\mathcal{P}.$ Hence, model approximations in this paper are focused on the dynamics of the hidden process. Henceforth, we will drop the superscript for emission densities. Note, also that the density 
	$\mathcal{P}_{\mathfrak{y}}$ appearing in equation (\ref{eq:dirac}) is entirely driven by $\mathcal{P}$ and binning. Therefore the emission densities do not involve any unknown parameters.
\end{remark}

A simple example is a sign classification of $\Delta y_{t+1},$ yielding a time series $\mathfrak{y}_{t}$ $\in$ $\{+, -\}$ with $n_O$ = 2. Of course finer bins, beyond the illustrative example of the sign indicator, will be more informative about the underlying return process.\footnote{In Definition \ref{def_obs} we assume that $S_{\Delta y}$ is closed and bounded, whereas the support of $\Delta y_t$ has in principle unbounded support, but it suffices to eliminate outcomes with arbitrary small probabilities.}
Next, we construct a latent discrete time and finite state Markov chain process of hidden states, namely:
\begin{definition}[Hidden states]
	\label{def_hidden}
	Define the set $\mathcal{H}_L$ := $\{h_1, \ldots, h_{n_L^c}\}$ of discrete hidden states where each $h_i$ is associated with a unique $V_i.$ Moreover, the dynamics of the hidden states are characterized by (a) a time homogeneous Markov chain with state transition matrix $\mathscr{A},$ and (b) a stochastic vector $\mathcal{X}_t \in [0, 1]^{n_L^c}$ representing the probability of being in state $i$ at time $t.$ 
	The number of hidden states $n_L^c$ is called the order of the HMM.
\end{definition}

\noindent Note that we used superscript $c$ for the number of hidden states but not for the observable ones, i.e.\ $n_O$ versus $n_L^c.$ The reason is that the classical HMM and the quantum counterpart will not have the same number of hidden states. Note also that we have not specified yet how $\bar{V}_t$ is driven by $V_t.$ This will be covered in the next subsection.\footnote{\label{foot:permutation} It should be noted that when we refer to the different hidden Markov models, it would be more appropriate to refer to them in terms equivalence classes since each model is specified up to a permutation of the hidden states. For convenience, we will keep permutation equivalence mostly out of the discussion.}

\medskip

The setup is known as a (classical) hidden Markov model (HMM) for a discrete time process with $n_L^c$ latent states and $n_O$ observed discrete-valued output states. More formally, it is defined as:
\begin{definition}[Classical Hidden Markov Model]
	\label{def_chmm}
	A classical hidden Markov model is defined by a 5-tuple: $\left(\mathcal{S}_O,\mathcal{H}_L, \mathscr{A}, \mathscr{E}, \mathcal{X}_0\right),$
	where $\mathcal{X}_0$ is a
	stochastic vector defining the initial latent process states and $\mathscr{E}$ the emission matrix. 
	The entires to $\mathscr{E}$ are determined by emission densities 
	$\mathcal{P}_{\mathfrak{y}}(\mathfrak{y}_{t+1} = \iota_i \vert V_j)$ appearing in equation 	(\ref{eq:dirac}) representing the probability of an observation in bin $i,$ given hidden state $V_j.$  A shorthand notation will be used for the classical HMM: $\mathscr{M}_{n_L^c}$ is the set of all classical HMM models with $n_L^c$ hidden states and a specific model of that type will be denoted $M_{n_L^c}$ without explicitly referring to the 5-tuple $\left(\mathcal{S}_O,\mathcal{H}_L, \mathscr{A}, \mathscr{E}, \mathcal{X}_0\right)$ associated with a specific model within that class. We will also write $\mathscr{A}(M_{n_L^c})$ since the transition density is model-driven.
\end{definition}

\noindent It is worth noting parenthetically that the evolution of the observation process, in general, is not  Markovian.  
To conclude, we also need to define two discrete time filtrations:   $\{\mathcal{F}_t^{\Delta y}, t \in T\}$ and $\{\mathcal{F}_t^{\mathfrak{y}}, t \in T\}.$ The former pertains to the $\sigma$-field filtration of past $\Delta y$ whereas the latter involves past $\mathfrak{y}.$

\subsection{Transition Kernels and Density Matrices \label{subsec:kernels}}

Recall that we will pursue two types of estimation approaches. Starting with the parametric approach, we need to specify functional forms for the drift $a(\cdot)$ and diffusion $b(\cdot)$ in equation (\ref{eq:SDE}). While there are no closed form transition kernels for SV models, there are cases where the transition kernel for the volatility equation is known. The most celebrated example is the so called Heston model with the square root or CIR process for $V_\tau,$ namely:
\begin{equation}
	\label{eq:Heston}
dV_\tau = \alpha(\beta - V_\tau) d \tau + \sigma \sqrt{V_\tau} dW_\tau,
\end{equation}
where $\alpha$ is the rate at which the process reverts to its long-term mean, equal $\beta,$ featuring a closed-form expression for the population transition kernel. In Appendix Section \ref{appsec:hestontrans} we present closed-form expression for the population transition kernel for a $\Delta/k$ increment, which is a non-central chi-squared distribution with $4\alpha \beta/\sigma^2$ degrees of freedom and non-centrality parameter $2 c V_\tau e^{-\alpha \Delta/k}$ with $c$ characterized in equation (\ref{appeq:transCIR1}). Moreover, the ergodic distribution of the process equals is a Gamma distribution which is also entirely determined by the three parameters driving the SDE in equation (\ref{eq:Heston}). 
Armed with the non-central $\chi^2$ distribution we can construct a discrete state Markov chain as follows: (a) we adopt a binning scheme similar to that for the observable returns and call it $\mathcal{B}_V$ := $\{b_1^V, \ldots, b_{n_L}^V\}$  (b)  we associate a spot volatility state $V_i$ with each bin to populate the set $\mathcal{H}_L$ of hidden states, each representing a midpoint of the binned partition, and finally construct the transition density matrix with transition probabilities from state $V_{t}$ = $V_i$ to $V_{t+1}$ = $V_j.$ 
In Appendix Section \ref{appsec:hestontrans} we detail these steps using a 4-bin example, where each bin occupies 25\% of the probability mass determined by the Gamma ergodic distribution. 

\medskip
 
When we don't want to commit to a diffusion equation for $V_t,$ we need to populate the transition density matrix $\mathscr{A}$ with $n_L^c(n_L^c-1)$ parameters 0 $<$ $a_{ij}$ $<$ 1 and $\sum_j a_{ij}$ = 1. This is obviously more costly particularly when $n_L^c$ gets large. This is the reason why we will consider penalized MLE with a penality a function of $n_L^c.$ Finally, as noted in Remark \ref{remark:discrSV}, our analysis covers discrete time SV models. In those cases the transition density is typically also easy to derive.

\subsection{Spot and Integrated Volatility \label{subsec:spotandint}}

The observable states are driven by  $\Delta y_t$ $\sim$ $\mathcal{N}(0,\bar{V}_t )$ conditional on $\mathfrak{U}_t.$ So far we only discussed the spot volatility process since the result in Lemma \ref{lemma:GJL} informed us that $\bar{V}_t$ is linked to $V_t.$ To establish such a link, we divide the discrete time interval $\Delta$ into $k$ equal disjoint increments. When we are working with a tightly parameterized diffusion, such as in Appendix equation (\ref{appeq:transCIR2}), we can compute for any 1 $\leq$ $j$ $\leq$ $k$ the transition probability $p(V_{t+j/k} | V_{t + (j-1)/k}).$ In case we use a fully parameterized density matrix $\mathscr{A},$ we can specify  $\mathscr{A}$ = $\mathscr{\tilde A}^k$ and work with $\mathscr{\tilde A}$ for the high-frequency transitions. In either case we rank the set of spot volatility hidden states  $\{h_1, \ldots, h_{n_L^c}\}$ in ascending order $V_1,$ $<$ $\ldots$ $<$ $V_{n_L^c}.$ Next, we compute realized volatility paths, ranked from low to high, starting with $kV_1,$ followed by $(k-1)V_1V_2,$ $\ldots$ $(k-1)V_{n_L^c}V_{n_L^c-1},$ and finally $kV_{n_L^c},$ for all paths of length $k.$ The number of distinct collections of length $k,$ which we shall denote by $n_{\bar{V}}^c$ is given by the formula: $n_{\bar{V}}^c$ = $\binom{n_L^c+k-1}{k}$ = $\frac{(n_L^c+k-1)!}{k!(n_L^c-1)!}.$
Standard HMM formulas allow us to attribute probabilities to each of these $n_{\bar{V}}^c$ paths, more on this in Section \ref{subsec:rank} equation (\ref{eq:classic_obs_prob}). Denoting said probability density as $g(\bar{V}_j \vert V_k; M_{n_L^c})$ for $j$ = 1, $\ldots,$ $n_{\bar{V}}^c$ and $k$ = 1, $\ldots,$ $n_L^c,$ for a given model $M_{n_L^c},$ the final step is to construct the emission matrix $\mathscr{E},$ using equation (\ref{eq:dirac}):
	\begin{equation}
	\label{eq:dirac}
	\mathcal{P}_{\mathfrak{y}}(\mathfrak{y}_{t+1} = \iota_i \vert \mathfrak{U}_t) = \sum_{j=1}^{n_{\bar{V}}^c}\int_{S_{\Delta y}} \mathcal{P}(x \vert \bar{V}_{t+1} = \bar{V}_j) g(\bar{V}_j \vert V_t; M_{n_L^c})\delta_{b_i}(x) d x \qquad \forall \iota_i \in \mathcal{S}_O, k = 1, \ldots, n_L^c.
\end{equation}
In this paper we will keep $k$ fixed, as it is beyond our ambition to endeavor into in-fill asymptotics. Therefore, there is only one source of increase in  $n_{\bar{V}}^c$ and that is $n_L^c,$ more specifically $n_{\bar{V}}^c$ is $O((n_L^c)^k).$ 

\subsection{Maximum Likelihood Estimation \label{subsec:MLEclassical}}

The parameters are collected in $\theta_c$ $\in$ $\Theta_c,$ a compact metric space. When we have a parametric diffusion in mind, as in equation (\ref{eq:Heston}), the parameter vector consists of $\alpha,$ $\beta$ and $\sigma.$ Similarly, if we work with discrete time SV models, as in Remark \ref{remark:discrSV}, we are also dealing with a few parameters describing the volatility dynamics (cfr.\ equation (\ref{eq:discreteSV})). If instead we select a fully parameterized transition density $\mathscr{A},$ than all its parameters appear in  $\theta_c.$  Note that in the former case, the dimension of the parameter space is independent of $n_L^c,$ whereas in the latter case the dimension grows with $n_L^c.$\footnote{The latter case is more common when discretizing continuous states, see e.g.\ \cite{bonhomme2022discretizing} for further discussion.} We treat  $\mathcal{X}_0$ as part of the parameters to estimate. Typically the Markov chain is initialized with its ergodic distribution, which itself depends on $\theta_c.$  For the different models we consider, the dimension of the parameter space is different. To avoid using $\theta$ as a vector whose dimension changes across models, and introducing convoluted notation, we will use $M_{n_L^c},$ suppressing the dependence on the parameter vector. 

\medskip

Traditional HMM models also have the emission matrix $\mathscr{E}$ as parameter-driven. Fortunately, this is not the case here as noted in Remark \ref{remark:emissiondensity}. Hence, increasing $n_O$ is of no consequence with respect to the dimension of model parameter space.\footnote{Of course, increasing $n_O$ does have computational implications which will be discussed later.}
Given a filtering algorithm providing the probability of being in state $i$ at time $t,$ i.e.\ $\mathcal{X}_{t-1}^i,$ we can write a log likelihood function for a sample $\mathfrak{Y}_{t=1}^\mathbf{T}$ := $\{\mathfrak{y}_t, t = 1, \ldots, \mathbf{T}\}$ under model $M_{n_L^c}^\mathfrak{y},$ associated with the  $\{\mathcal{F}_t^{\mathfrak{y}}, t \in T\}$ filtration, as:
\begin{eqnarray}
	\label{eq:log-likeBiny}
	\mathscr{L}(\mathfrak{Y}_{t=1}^\mathbf{T}; M_{n_L^c}^\mathfrak{y}) & := & \sum_{t=1}^{\mathbf{T}}  \ell_t^{\mathfrak{y}}(M_{n_L^c}^\mathfrak{y}, \mathcal{X}_{t-1}) \\
		\ell_t^{\mathfrak{y}}(M_{n_L^c}^\mathfrak{y},\mathcal{X}_{t-1}) & := &  \sum_{i=1}^{n_L^c}\sum_{j=1}^{n_{\bar{V}}^c}\log \mathcal{P}_{\mathfrak{y}}(\mathfrak{y}_{t} \vert \mathfrak{U}_t) g(\bar{V}_{t-1} = \bar{V}_j \vert V_{t-1} =  V_i; M_{n_L^c}^\mathfrak{y})\mathcal{X}_{t-1}^i. \notag
\end{eqnarray}
Next we consider the $\{\mathcal{F}_t^{\Delta y}, t \in T\}$ filtration, yielding the log likelihood for a sample $\Delta \mathscr{Y}_{t=1}^\mathbf{T}$ := $\{\Delta y_t, t = 1, \ldots, \mathbf{T}\}$ under model $M_{n_L^c}^{\Delta y},$ namely:
\begin{eqnarray}
	\label{eq:log-likeDeltay}
		\mathscr{L}(\Delta \mathscr{Y}_{t=1}^\mathbf{T}; M_{n_L^c}^{\Delta y}) & := & \sum_{t=1}^{\mathbf{T}}  \ell_t^{\Delta y}(M_{n_L^c}^{\Delta y}, \mathcal{X}_{t-1}) \\
	\ell_t^{\Delta y}(M_{n_L^c}^{\Delta y},\mathcal{X}_{t-1}) & := &    \sum_{i=1}^{n_L^c}\sum_{j=1}^{n_{\bar{V}}^c}\left[-\frac{1}{2} \ln(2\pi \bar{V}_{j}) - \frac{(\Delta y_t)^2}{2\bar{V}_{j}}\right] g(\bar{V}_{t-1} = \bar{V}_j \vert V_{t-1} =  V_i; M_{n_L^c}^{\Delta y})\mathcal{X}_{t-1}^i. \notag
\end{eqnarray}
To compute the likelihood we also need to characterize the filtering algorithm for $\mathcal{X}_{t}.$ The filtering algorithm applies, mutatis mutandis, to both filtrations $\{\mathcal{F}_t^{\mathfrak{y}}, t \in T\}$ and $\{\mathcal{F}_t^{\Delta y}, t \in T\}.$ Namely, following \cite{hamilton2020time}, pp.\ 692-693, making explicit the dependence of the transition density matrix and emission on model $M_{n_L^c}$ (either $M_{n_L^c}^\mathfrak{y}$ or $M_{n_L^c}^{\Delta y}$) and using the Hadamard element-wise product notation $\odot$ we have:
\begin{equation}
	\label{eq:fliterclassical}
	\mathcal{X}_t = \frac{\mathscr{A}(M_{n_L^c}) \mathcal{X}_{t-1} \odot \mathscr{E}_t(M_{n_L^c})}{{\bf{1}^\prime} \left[\mathscr{A}(M_{n_L^c}) \mathcal{X}_{t-1} \odot \mathscr{E}_t(M_{n_L^c})\right]}
\end{equation} 
where $\bf{1}$ is a $n_L^c$ vector of ones, and $\mathscr{E}_t(M_{n_L^c})$ := $\mathcal{P}_{\mathfrak{y}}(\mathfrak{y}_{t} \vert \mathfrak{U}_t) g(\bar{V}_{t-1} = \bar{V}_j \vert V_{t-1} =  V_i; M_{n_L^c})$ for the $\mathcal{F}_t^{\mathfrak{y}}$ filtration and $\mathscr{E}_t$ := $\mathcal{P}_{\Delta y}(\Delta y_{t} \vert \mathfrak{U}_t) g(\bar{V}_{t-1} = \bar{V}_j \vert V_{t-1} =  V_i; M_{n_L^c})$ for the $\mathcal{F}_t^{\Delta y}$ filtration. While the two filtrations have the same updating formula, they differ in sample paths of the realized filters. We will therefore refer to $\mathcal{X}_t^\mathfrak{y}$ for the $\mathcal{F}_t^{\mathfrak{y}}$ filtration and $\mathcal{X}_t^{\Delta y}$ for $\mathcal{F}_t^{\Delta y}.$ 



Finally, since we suppressed explicit mention of parameters we will denote the maximum likelihood estimates for model $M_{n_L^c}$ by $\hat M_{n_L^c}.$

\subsection{Rank of HMM \label{subsec:rank}}

To conclude we discuss a notion of rank introduced  by \cite{anderson1999realization} using an infinite generalized Hankel matrix representation of a HMM. To construct such matrix we consider $\mathfrak{Y}_{t_1}^{t_2}$ for finite $t_i$ $\in $ $T$ consisting of a concatenation of $\mathfrak{y}_{t_1}, \ldots, \mathfrak{y}_{t_2}.$ In addition, we also define the set $\mathcal{Y}_{\mathfrak{y}}$ :=  $\{\emptyset, \mathfrak{Y}_{t_1}^{t_2} ,\ \forall t_1, t_2 \in T \text{ s.t. } t_2 - t_1 \text{ finite}\}.$ 
For a single event, we have $\mathcal{P}_{\mathfrak{y}}(\mathfrak{y}_{\mathbf{T}+1} = \iota_i \vert \mathfrak{U}_{\mathbf{T}})$ appearing in equation (\ref{eq:dirac}), and for a string of events we can  define a set of operators $\mathbf{T}$ as follows
\begin{equation}
	\label{eq:classicobsop}
	\mathbf{T} = \{ \mathbf{T}_{\iota_i} : \mathbf{T}_{\iota_i}= \mathscr{A} \mathscr{E}_{\iota_i}, \iota_i \in \mathcal{S}_O  \},
\end{equation}
where $\mathscr{E}_{\iota_i}$ = diag($\mathscr{E}[\iota_i,h_j]$), $\iota_i$ $\in$ $\mathcal{S}_O$ are diagonal matrices with
$h_j$ $\in$ $\mathcal{H}_L$ and $\mathscr{A}$ is the transition matrix with $\mathscr{A}$ = $\sum_{\iota_i \in \mathcal{S}_O} \mathbf{T}_{\iota_i}.$
Let $\{\mathfrak{y}_{t_1} = \iota_{t_1}, \ldots, \mathfrak{y}_{t_2} = \iota_{t_2} \}$ be any finite sequence of observations, then:
\begin{equation}
	\label{eq:classic_obs_prob}  
	\mathcal{P}[\mathfrak{y}_{t_1} = \iota_{t_1}, \ldots, \mathfrak{y}_{t_2} = \iota_{t_2}]=\textbf{1}\mathbf{T}_{\iota_{t_2}} \ldots \mathbf{T}_{\iota_{t_1}}\mathcal{X}_0,
\end{equation}
where \textbf{1} is the unit row vector with dimension $n_O.$
Following Assumption \ref{assum:naturalorder} we have a natural order for the bins, which allows us to use a lexicographical order of observations for any finite sequence in $\mathcal{Y}_{\mathfrak{y}},$ and construct the following bi-infinite Hankel matrix (also known as a generalized Hankel matrix):
\begin{equation}
	\label{eq:hankel}
	\mathcal{H}_{HMM} := 
	\left[
	\begin{array}{cccccccc}
		\mathcal{P}(\emptyset) & \mathcal{P}(\iota_1) & \mathcal{P}(\iota_2) & \cdots & \mathcal{P}(\iota_{n_O})  & \mathcal{P}(\iota_1 \iota_1) & \mathcal{P}(\iota_1 \iota_2) & \cdots \\
		\mathcal{P}(\iota_1) & \mathcal{P}(\iota_2) & \cdots & \mathcal{P}(\iota_{n_O})  & \mathcal{P}(\iota_1 \iota_1) & \mathcal{P}(\iota_1 \iota_2) & \mathcal{P}(\iota_1 \iota_3) & \cdots \\
		\vdots & \ddots &  & \vdots &   & \vdots & & \cdots 
	\end{array}
	\right]
\end{equation}

\begin{definition}[Rank of HMM]
	\label{def:rankHMM}
If a HMM has an infinite generalized Hankel matrix of rank $n,$ then rank(HMM) $\leq n$.
\end{definition} 
Likewise, the rank of a stochastic process is the rank of its Hankel matrix according to the definition of \cite{anderson1999realization}. This matrix is important, because its finite-rank property is a necessary condition for the stochastic process to have a HMM realization. Since we will think of hidden Markov models as approximations, it is not key for our analysis to assume that the stochastic process has indeed a finite rank. What will matter is the rank of the HMM we select in order to compare the classical HMM with the quantum one.

\medskip

Note that in Definition \ref{def_hidden} the order of the HMM is equal to the number of hidden states. In \cite{huang} it is shown the inequality in Definition \ref{def:rankHMM} can be reached, i.e.\ there is a minimal HMM with rank equal to its order. In the remainder of the paper we will maintain Assumption \ref{assum:orderrank} which states that the HMM is minimal, i.e.\ rank and order are equal.

\medskip

The (engineering) literature on HMM makes a difference between the so called Moore HMM and Mealy HMM. Although there is a correspondence between the two, we work with the former. It was noted in footnote \ref{foot:permutation} that it would be more appropriate to refer to such models in terms equivalence classes since each model is specified up to a permutation of the hidden states. Fortunately, the minimal Moore representation is unique (under suitable regularity conditions applicable to our case) except for the trivial permutations of the hidden states, see \cite{vanluyten2008equivalence}, in particular their Theorem 4, for further discussion.

\setcounter{equation}{0}
\section{Diffusions and Quantum Hidden Markov Models \label{sec:QHMM}}

Quantum computing (QC) centers around the harnessing of quantum physical phenomena, such as superposition and entanglement, to perform computations -- see \cite{nielsen_chuang_2010}. In this section, we focus on how to formulate a Quantum HMM (henceforth QHMM). To do so, we need to define some key quantum concepts pertaining to states, state evolution, the measurement process and observables. The first feature of the QHMM distinct from the classical HMM is that we are dealing with quantum states.  For such states we adopt the  \cite{dirac1939new} bra-ket notation used in quantum mechanics, as well as the quantum computation and information literature.\footnote{See \cite{nielsen_chuang_2010} for a standard textbook reference or the Appendix to \cite{morrell2021step} which provides a quick summary focused on the concepts used in our paper .In Appendix section \ref{appsec:QC} we review some quantum information basic concepts used in the formulation of the QHMM. The Dirac bra-ket notation uses $\ket{u_i}$ (called a ket) for the complex column vector $u_i$ and $\bra{u_i}$ (a bra) is written for its adjoint, the row vector containing the complex conjugates of its elements, with the complex conjugate written as $\overline{u}_i$. The quadratic form $\overline{u}_i A u_i$ is then written as $\bra{u_i} A \ket{u_i}$. The notation allows one to distinguish numbers from matrices, as in $\braket{u_i}{u_i}$ versus $\ket{u_i} \bra{u_i}$, and to specify vectors through labels or descriptions as in $\ket{\text{Model}\, 1}$.} 

\medskip

Our goal is to formulate a quantum counterpart to $\mathscr{L}(\mathfrak{Y}_{t=1}^\mathbf{T}; M_{n_L^c}^\mathfrak{y}),$ the likelihood function  appearing in equation (\ref{eq:log-likeBiny}). 
From Definition \ref{def_obs} we know that $\{\mathfrak{y}_{t}: t \in T \subset \mathbb{N}, \Delta y_t \in b_i \text{ then } \, \mathfrak{y}_{t} = \iota_i \}$ where $\iota_i$ is the $i^{th}$ coordinate vector of $\mathbb{R}^{n_o}.$ To formulate a quantum counterpart, we must first recognize that the mathematical formalism of QC involves Hilbert spaces. Therefore, instead of using the $i^{th}$ coordinate vector of $\mathbb{R}^{n_o}$ to label $\mathfrak{y}_{t}$ observations, we will use an isomorphic orthonormal basis $\ket{u_i^O}$ for $i$ = 0, $\ldots$, $n_O-1$ of a tensor product space $\hilbert_O$ := $\bigotimes_{i=1}^{n_O} \hilbert$ to formulate the likelihood of a sample $\mathfrak{Y}_{t=1}^\mathbf{T}.$ The model that generates these likelihoods will be different, and represented as $\mathscr{L}(\mathfrak{Y}_{t=1}^\mathbf{T}; M_{n_L^q}),$ where  $M_{n_L^q}$ is a shorthand for a QHMM with $n_L^q$ hidden states (since QHMM only applies to $\mathfrak{y}$ we drop the superscript to the model labels).
Not surprisingly, there is no unique way to define a QHMM. Some characterizations are observationally equivalent, while others define potentially different stochastic processes.\footnote{See for example \cite{monras2011hidden}, \cite{adhikary2020expressiveness} and \cite{markov2022implementation} for further discussion.} The definition we adopt is one amenable to quantum computational implementation put forward by \cite{markov2022implementation}, which they call the unitary Quantum Hidden Markov Model. Since we will only consider this particular formulation we will simply call it the QHMM, dropping the unitary label.

\medskip

Mathematically, a QHMM is an example of what is called a quantum channel, which describes how a composite quantum system evolves according to its internal dynamics while simultaneously parts of it are observed by measurement. The model combines quantum unitary hidden states evolution with observable states correlated with the hidden states. More specifically, the QHMM formulation we adopt is as follows:

\begin{definition}[Unitary Quantum Hidden Markov Model]
	\label{def_unitary_qhmm}
	A unitary Quantum HMM is defined by a 6-tuple: $\left(\mathcal{S}_O,\hilbert_L, \hilbert_O, U, \mathcal{M}, R_0\right),$
	where: (a) $\mathcal{S}_O$ as in Definition \ref{def_obs}. Hence, output states in the classical and quantum HMM are the same, (b) $\hilbert_L$  the tensor product space $\hilbert_L$ := $\bigotimes_{i=1}^{n_L^q} \hilbert$ associated with hidden quantum states and orthonormal basis $\ket{u_i^L}$ for $i$ = 0, $\ldots$, $n_L^q-1,$ where $n_L^q$ refers to the number of hidden quantum states, (c) $\hilbert_O$  the tensor product space $\hilbert_O$ := $\bigotimes_{i=1}^{n_O} \hilbert$ associated with the observable states and orthonormal basis $\ket{u_i^O}$ for $i$ = 0, $\ldots$, $n_O-1$ with $n_O$ $\leq$ $(n_L^q)^2,$ (d) $U$ is a unitary operator defined on the bipartite Hilbert space $\hilbert_L \otimes \hilbert_O,$ (e) $\mathcal{M}$ is a bijective map $\mathcal{P}_i^{O} \rightarrow \mathcal{S}_O$, where $\mathcal{P}_i^{O}$ is a partition of the orthonormal basis $\ket{u_i^O}$ for $i$ = 0, $\ldots$, $n_O-1$ and finally (f) $ R_0 = \rho_0^L \otimes \ket{u_0}\bra{u_0}$ is an initial emission pure state where $\ket{u_0}$ $\in$ $\{\ket{u_i^O}\}$ and $\rho_0$ is the initial latent mixed state, i.e.\ $\rho_0^L$ $\in$  $\mathcal{D}(\hilbert_L)$ the space of density operators on $\hilbert_L.$
\end{definition}

Similar to the definition of classical HMM, we'll use a shorthand notation for the quantum HMM: $\mathscr{M}_{n_L^q}$ is the set of all QHMM models with $n_L^q$ hidden states and a specific model of that type will be denoted $M_{n_L^q}$ without  referring explicitly to the 6-tuple: $\left(\mathcal{S}_O,\hilbert_L, \hilbert_O, U, \mathcal{M}, R_0\right).$  We will also write $U(M_{n_L^q})$ since the transition density is model-driven.

To represent a quantum system composed of two subsystems, we use a quantum operation that combines the components through a tensor product, creating a bipartite quantum system: $\rho_L$ $\mapsto$ $\rho_L \otimes \rho_O$ := $\rho_{LO},$ represented by the Hilbert space $\hilbert_{LO}$ = $\hilbert_L \otimes \hilbert_O,$ where $\rho_{LO}$ is a tensor product state where the subsystems are not entangled. 
To model the joint stochastic processes of $\{\mathfrak{y}_t, \mathfrak{U}_t\}$ in a quantum setting, or put differently, the joint evolution of quantum states in $\hilbert_L$ and in $\hilbert_O,$ the unitary  operator $U$ is applied to entangle the subsystems and transfer information from the hidden state to the observable ones.
The evolution of the hidden state is then described by tracing out the $O$ component from the entangled system. The technical details of this appear in Appendix Section \ref{appsec:composite}. 

\medskip

The QHMM involves parameters that are collected in $\theta_q$ $\in$ $\Theta_q,$ a compact metric space. Note the subscripts $q$ to distinguish these from the classical HMM parameters. To streamline notation again, we will use $M_{n_L^q},$ representing the model, instead of the parameter vector when writing the log likelihood functions.
The log likelihood function can then be written as:
{\small
\begin{equation}
	\label{eq:log-likeBinyQ}
	\mathscr{L}(\mathfrak{Y}_{t=1}^\mathbf{T}; M_{n_L^q}) := 	\ln \text{tr} M_{\mathbf{T}}\left(\sum_i{K_i{\dots   M_{2}\left(\sum_i{K_i{M_{1}\left(\sum_i{K_i\rho_0^LK_i^\dagger}\right) M_{1}^\dagger}K_i^\dagger}\right) M_{2}^\dagger\dots}K_i^\dagger}\right) M_{\mathbf{T}}^\dagger
\end{equation}}
associated with the  $\{\mathcal{F}_t^{\mathfrak{y}}, t \in T\}$ filtration and where $K_i$ are known as Kraus operators. Here they depend on the model parameters, namely:
\begin{equation}
	\label{eq:kraus_op}
	K_i = \left( I_{n^q_L} \otimes \bra{u_i^O} \right) U(\theta_q) \left( I_{n^q_L} \otimes \ket{u_0} \right),
\end{equation}
and the filtering algorithm for the latent quantum states appears in Appendix equation (\ref{appeq:quantfilter}).

\medskip

To conclude we revisit the rank of HMM in the context of a quantum setting. 
Using the results in Appendix Section \ref{appsec:composite} we have for a string of observations $\mathfrak{Y}_{t_1}^{t_2}:$  $\{\mathfrak{y}_{t_1} = \iota_{t_1}, \ldots, \mathfrak{y}_{t_2} = \iota_{t_2} \}:$ 
\begin{equation}
	\label{eq:evolution_measurement_multi}
		\mathcal{P}[\mathfrak{y}_{t_1} = \iota_{t_1}, \ldots, \mathfrak{y}_{t_2}] = 
		\quad \text{tr}\left\{M_{t_2}\left(\sum_i{K_i{\dots \left(\sum_i{K_i{M_{t_1}\left(\sum_i{K_i\rho_0^LK_i^\dagger}\right) M_{t_1}^\dagger}K_i^\dagger}\right) \dots}K_i^\dagger}\right) M_{t_2}^\dagger\right\}.
\end{equation}
With these probabilities we can populate the bi-infinite Hankel matrix appearing in equation (\ref{eq:hankel}). Note that we have both the classical and quantum HMM providing comparable entries to the Hankel matrix and therefore allows us study their rank properties, among other things.

\setcounter{equation}{0}
\section{Approximation and Quantum Dimensionality Reduction \label{sec:dimapprox}}

Given the likelihood under the true DGP, denoted by $\mathscr{L}^*$ - which is well defined, see proof of Lemma \ref{lemma:approxbins} for further discussion - we define two Kullback-Leibler (KL) divergence measures using the likelihood functions appearing in (\ref{eq:log-likeBiny}) and (\ref{eq:log-likeDeltay}):
\begin{equation}
	\label{eq:KL1}
	D_{KL}(\mathscr{L}^*_{\mathbf{T}} \parallel \mathscr{L}_{\mathbf{T}}^{\mathfrak{y}}(M_{n^c_L}^\mathfrak{y})) = \mathbb{E}^*\left[\log \frac{\mathscr{L}^*( \Delta \mathscr{Y}_{t=1}^\mathbf{T})}{\mathscr{L}( \mathfrak{Y}_{t=1}^\mathbf{T}, M_{n^c_L}^\mathfrak{y})}\right],
\end{equation}
and:
\begin{equation}
	\label{eq:KL2}
	D_{KL}(\mathscr{L}^*_{\mathbf{T}} \parallel \mathscr{L}_{\mathbf{T}}^{\Delta y}(M_{n^c_L}^{\Delta y})) = \mathbb{E}^*\left[\log \frac{\mathscr{L}^*( \Delta \mathscr{Y}_{t=1}^\mathbf{T})}{	\mathscr{L}(\Delta \mathscr{Y}_{t=1}^\mathbf{T}, M_{n^c_L}^{\Delta y})}\right],
\end{equation}
where $\mathbb{E}^*[.]$ is the expectation under the true DGP. We also define:
\begin{equation*}
	D_{KL}(\mathscr{L}_{\mathbf{T}}^{\Delta y}(M_{n^c_L}^{\Delta y})\parallel  \mathscr{L}( \mathfrak{Y}_{t=1}^\mathbf{T}, M_{n^c_L}^\mathfrak{y})) :=  \mathbb{E}^*\left[\log \frac{\mathscr{L}(\Delta \mathscr{Y}_{t=1}^\mathbf{T}, M_{n^c_L}^{\Delta y})}{ \mathscr{L}( \mathfrak{Y}_{t=1}^\mathbf{T}, M_{n^c_L}^\mathfrak{y})}\right].
\end{equation*}
Lemma \ref{lemma:approxbins} in the Appendix shows that  there is $n_O$ such that $D_{KL}(\mathscr{L}_{\mathbf{T}}^{\Delta y}(M_{n^c_L}^{\Delta y})\parallel  \mathscr{L}( \mathfrak{Y}_{t=1}^\mathbf{T}, M_{n^c_L}^\mathfrak{y}))$
$<$ $\eta$ for any given $\eta$ $>$ 0, and $\mathbf{T}.$ For this reason we will focus on the likelihood functions involving the discrete time filtration  $\{\mathcal{F}_t^{\mathfrak{y}}, t \in T\},$ and assume that it arbitrarily close to the continues observation process $\Delta y.$\footnote{Note that this is a statistical argument, not a computational one. Indeed, recall that the emission schemes in our case do not involve any additional parameters, and therefore increasing $n_O$ is  solely a refinement of the information sets, although one has to keep in mind that finer bins of discretizations entail additional computational costs.} Moreover, we will drop the superscripts referring to the data filtrations and write: $\mathscr{L}_{\mathbf{T}}(M_{n^c_L}).$ Finally,
 for the QHMM we also have:
\begin{equation}
	\label{eq:KL3}
	D_{KL}(\mathscr{L}^*_{\mathbf{T}} \parallel \mathscr{L}_{\mathbf{T}}^{\mathfrak{y}}(M_{n^q_L}^\mathfrak{y})) = \mathbb{E}^*\left[\log \frac{\mathscr{L}^*( \Delta \mathscr{Y}_{t=1}^\mathbf{T})}{\mathscr{L}( \mathfrak{Y}_{t=1}^\mathbf{T}, M_{n^q_L}^\mathfrak{y})}\right].
\end{equation}

\medskip

\begin{remark}
	\label{remark:shaodow}
	Although the classical likelihood ratio in quantum
	system estimation can be naturally introduced as the ratio of  probabilities of obtaining a measurement outcome on
	two different quantum states, the ``quantum likelihood ratio'' does not yet have a confirmed definition (see \cite{yamagata2013quantum} for further discussion) - and by the same token a likelihood ratio of classical/quantum models. Following \cite{yano2024statistical}, among others, we rely on the notion of classical shadows developed by \cite{huang2020predicting} to have a classical representation of the log-likelihood ratio. 
\end{remark}

We want to compare classical with quantum HMM through their respective sample likelihood functions: $\mathscr{L}_{\mathbf{T}}(M_{n^c_L})$ and $\mathscr{L}_{\mathbf{T}}(M_{n^q_L})$ focusing on the key model feature $n_L^c$ versus $n_L^q.$  

\begin{assumption}
	\label{assum:LLN}
	We assume the following laws of large numbers apply as $\mathbf{T}$ $\rightarrow$ $\infty:$
	\begin{equation}
		\label{eq:LLN1}
		D_{KL}(\mathscr{L}^*_{\mathbf{T}} \parallel \mathscr{L}_{\mathbf{T}}(\hat M_{n^c_L}))  \underset{p}{\rightarrow}  	\underset{M_{n_L^c} \in \mathscr{M}_{n_L^c}}{\text{Argmin}}	D_{KL}(\mathscr{L}^*_{\mathbf{T}} \parallel \mathscr{L}_{\mathbf{T}}( M_{n^c_L})) 
 := D_{KL}^*(M_{n_L^c}^{*}) 
	\end{equation}
where $\mathscr{M}_{n_L^c}$ is the set of all classical HMM models with $n_L^c$ hidden states and $M_{n_L^c}^{*}$ is the model within that class that minimizes the KL-divergence. Similarly, for the quantum HMM and $\mathscr{M}_{n_L^q}$ the set of all QHMM models with $n_L^q$ hidden states :
	\begin{equation}
	\label{eq:LLN2}
	D_{KL}(\mathscr{L}^*_{\mathbf{T}} \parallel \mathscr{L}_{\mathbf{T}}(\hat M_{n^q_L}))  \underset{p}{\rightarrow}  	\underset{M_{n_L^q} \in \mathscr{M}_{n_L^q}}{\text{Argmin}}	D_{KL}(\mathscr{L}^*_{\mathbf{T}} \parallel \mathscr{L}_{\mathbf{T}}( M_{n^q_L})) 
	:= D_{KL}^*(M_{n_L^q}^{*}) 
\end{equation}
\end{assumption}
\noindent The regularity conditions pertaining to the convergence in probability appearing in equation (\ref{eq:LLN1}) are covered in the existing asymptotic theory for classical HMM models, see e.g.\ \cite{mevel2004asymptotical}. In contrast, the asymptotic theory regarding quantum HMM, assuming the latter is an approximate model to the true DGP, is not well developed as further discussed later. 
Next, we introduce the notion of KL-equidivergence:.

\begin{definition}[KL-equidivergence]
	\label{def:KL-equiv}
	Two hidden Markov models, $M_1$ and $M_2,$ are KL-equidivergent if and only if $D_{KL}^*(M_1)$ = $D_{KL}^*(M_2),$ assuming both are well defined, see Assumption \ref{assum:LLN}. 
\end{definition}
\noindent The notion of KL-equidivergence leads to the following theoretical result:

\begin{thm}
	\label{theorem:quadraticdimred}
	Let Assumptions \ref{assum:knownhidden} through 
		\ref{assum:orderrank} and \ref{assum:LLN} hold. Then using Definition \ref{def:KL-equiv},
	for every classical HMM with $n_L^c$ = $n^2$ hidden states there is at least one KL-equidivergent quantum HMM with $n_L^q$ = $n$ hidden states. Let $\mathcal{M}^{q}_{\sqrt{n^c_L},\mathbf{T}}(M)$ be the set of quantum HMM models, with $\sqrt{n^c_L}$ hidden states, KL-equidivergent to the classical HMM model $M$ $\in$ $\mathcal{M}_{n^c_L,\mathbf{T}}.$
\end{thm}

The result tells us that there is a quadratic improvement in the dimensionality reduction of HMM. Quantum computing algorithms can offer various types of speedups over classical algorithms. Quadratic speedups are achieved by Grover's algorithm, quantum amplitude estimation, Monte Carlo simulation-based estimation, whereas polynomial speedups are achieved by the HHL algorithm (\cite{harrow2009quantum}) to solve systems of linear equations (see \cite{ghysels2024quantum} and \cite{morgan2025enhanced} for asset pricing implications), quantum approximate optimization algorithm (QAOA), quantum phase estimation for eigenvalue estimation, quantum principal component analysis, among others.\footnote{Broadly speaking, quantum computation holds transformative potential for the finance industry, offering solutions to complex problems that are computationally infeasible for classical computers.  
	See \cite{orus2019quantum} and \cite{herman2022survey} for recent surveys.} Theorem \ref{theorem:quadraticdimred} adds KL-equidivergence to the list of quadratic speedups.\footnote{It should parenthetically be noted that the set of quantum HMM models, with $\sqrt{n^c_L}$ hidden states, KL-equidivergent to the classical HMM model in Theorem \ref{theorem:quadraticdimred} is not a singleton, since as noted in footnote \ref{foot:permutation} each model represents an equivalence class due to permutations of hidden states.}  The next theorem will add quadratic improvements in likelihood ratios.
\begin{thm}
	\label{theorem:quadraticLR}
	Let Assumptions \ref{assum:knownhidden} through 
	\ref{assum:orderrank} and \ref{assum:LLN} hold. 
	Consider a classical HMM with $n_L^c$ = $n^2$ hidden states and quantum HMM with $n_L^q$ = $n$ hidden states. Then for $n_L^q$ = $\sqrt{n^c_L},$ and as $\mathbf{T}$ $\rightarrow$ $\infty:$ $\mathbb{E}^*\left[\log (\mathscr{L}(\Delta \mathscr{Y}_{t=1}^\mathbf{T}; M_{n^q_L})/	\mathscr{L}(\Delta \mathscr{Y}_{t=1}^\mathbf{T}; \tilde M_{n^c_L}))\right]$ $\geq$ 0,
or equivalently:
$D_{KL}^*(M_{n^q_L})$ $\leq$ $D_{KL}^*(\tilde{M}_{n^c_L}).$
\end{thm}

Note that, while the asymptotic theory for approximate quantum HMM is still challenging as discussed later, we expect even in the classical case a non-standard asymptotic distribution for the log likelihood ratio appearing the in above theorem. Indeed, the fact that there is a reduction of the number of hidden state results in parameter boundary restrictions which yields non-standard large sample theory (see e.g.\ \cite{garcia1998asymptotic} and \cite{carrasco2014optimal} for further discussion). A priori we don't expect the log likelihood ratio to be positive, but Theorem \ref{theorem:quadraticLR} tells us that in large samples this is the case for the quantum versus classical HMM reduced hidden states comparison.

\section{Empirical Implementation \label{sec:implementation}}

The goal of this section is  (a) to provide an example of comparable classical and quantum HMMs, and (b) to see the practical benefits of the theory presented in Section \ref{sec:dimapprox}. Our analysis is a proof of concept rather than a comprehensive empirical study.
We estimate a tightly parameterized classical HMM using daily log returns of the S\&P 500 from 1990-2024. In particular, we estimate a CIR-type parametric classical with 16 hidden states, as detailed in Appendix Section \ref{appsec:implementation}. The empirical model will be used as our data generating process for simulations comparing classical and  quantum HMM. 

\medskip

In subsection \ref{sec:im_qhmm} we describe the specifics of the QHMM we use in the simulations. Next, we cover hardware implementation issues as they pertain to the Markov properties of QHMM ins subsection \ref{subsec:Markov} followed by a discussion of complexity analysis in subsection \ref{subsec:complexity}. The final subsection covers empirical findings.

\subsection{QHMM Model Specification}
\label{sec:im_qhmm}

The definition in Section \ref{sec:QHMM} assumes a unitary QHMM that consists of any quantum gate, and an observation in any basis. Here we outline the Ansatz circuit implementation with observations in the computational basis inspired by \cite{markov2022implementation}.\footnote{In quantum computing, the term "Ansatz" refers to a trial wave function or trial state used as a starting point for approximations or optimizations.} We choose this approach because it is realistic to a Quantum Processor (QPU), is consistent with trends in other areas of quantum machine learning (see e.g.\ \cite{Schuld:2021mml}), and can be estimated with the same numerical methods as the classical HMMs.\footnote{Current QPUs are not able to implement a QHMM for a realistic number of time steps, and therefore we classically calculate the likelihood function in equation (\ref{eq:log-likeBinyQ}) for each set of parameters with Qiskit software, see \cite{javadiabhari2024quantumcomputingqiskit}. Qiskit is the world’s most popular open source software stack for quantum computing.} 

For a fair comparison, we want a QHMM framework that can be implemented with the same estimation algorithm as the classical HMM. For this reason we use an Ansatz circuit to structure our unitary QHMM. The primary benefit of this approach is that it leaves us with a constant number of parameters that can be used to construct a given model similar to the classical counterparts. Keeping the circuit constrained to a fixed structure provides practical benefits for implementation and enables us to perform more accurate resource estimates.  When selecting an Ansatz circuit we considered common options such as an EfficientSU2 or RealAmplitudes circuits that are standard features of Qiskit's circuit library. Deeper Ansatz circuits are more expressive, meaning they can recreate a larger portion of the total possible QHMMs. This expressiveness comes at the cost of added computational complexity and additional parameters to optimize. We found an EfficientSU2 circuit with full entanglement and three rotation layers to be a reasonable compromise between these two factors. While current hardware is not yet ready to implement a QHMM with a realistic amount of data, using an Ansatz circuit assures that our method will be easily transferable to hardware once sufficient QPUs are available. 

\medskip

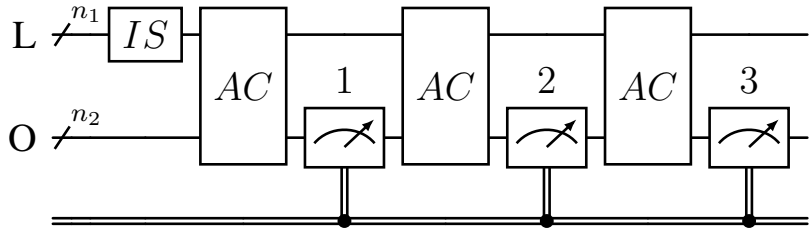
\begin{figure}
	\centering
	\begin{tikzpicture}
		\node[scale=.8](circ1){		
			\begin{adjustbox}{width=0.8\columnwidth}
				\begin{quantikz}[column sep=0.2cm, wire types={q, q, c}]
					\lstick{L} & \qwbundle{n_1} & & \gate{IS} & \gate[2]{AC} & & \gate[2]{AC} & & \gate[2]{AC} & &\\
					\lstick{O} & \qwbundle{n_2} & & & & \meter{1} \wire[d][1]{c} & & \meter{2} \wire[d][1]{c} & & \meter{3} \wire[d][1]{c} &  \\
					& & & & & \control{} & & \control{} & & \control{} &
				\end{quantikz}
			\end{adjustbox}
		};
	\end{tikzpicture}
	\caption{Three time step generic Ansatz circuit QHMM. The latent register ``L'' contains $n_1 = \log_2(n^q_L)$ qubits. The observed register ``O'' contains $n_2 = \log_2(n_O)$ qubits. The ``IS'' gate can be any gate chosen to prepare the initial gate of the latent system. The ``AC'' gate is the chosen Ansatz circuit. The observed register is reset to the $\ket{0}$ state after each measurement. \label{fig:circ_qhmm}}
\end{figure}

\begin{figure}
	\centering
	\begin{tikzpicture}
		\node[scale=.75](circ1){		
			\begin{adjustbox}{width=0.8\columnwidth}
				\begin{quantikz}[column sep=0.2cm, wire types={q, q, q, q, c}]
					\lstick{$\ket{0}_{L1}$} & \gate{H} & \ctrl{1} & & \gate{R_y} & \gate{R_z} & \ctrl{1} & \ctrl{2} & & \ctrl{3} & & & \gate{R_y} & \gate{R_z} & & \\
					\lstick{$\ket{0}_{L2}$} & & \targ{} & & \gate{R_y} & \gate{R_z} & \targ{} & & \ctrl{1} & & \ctrl{2} & & \gate{R_y} & \gate{R_z} & & \\
					\lstick{$\ket{0}_{O1}$} & & & & \gate{R_y} & \gate{R_z} & & \targ{} & \targ{} & & & \ctrl{1} & \gate{R_y} & \gate{R_z} & \meter{} \wire[d][2]{c} & \\
					\lstick{$\ket{0}_{O1}$} & & & & \gate{R_y} & \gate{R_z} & & & & \targ{} & \targ{} & \targ{} & \gate{R_y} & \gate{R_z} & & \meter{} \wire[d][1]{c} \\
					& & & & & & & & & & & & & & \control{} & \control{}
				\end{quantikz}
			\end{adjustbox}
		};
	\end{tikzpicture}
	\caption{Example circuit of QMM using an EfficientSU2 Ansatz circuit for one time step. We use $Ry$ and $Rz$ rotation gates with one repetition. In our training process, we use 3 repetitions which results in a single Ansatz circuit depth of 21 gates.\label{fig:esu2}}
\end{figure}
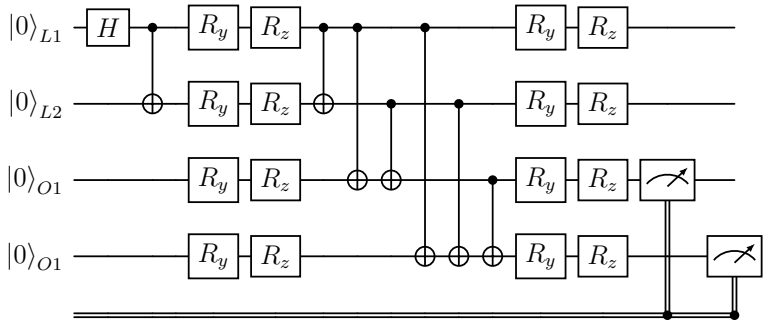

In the general case, a unitary QHMM circuit consists of two quantum registers. The latent register needs $\log_2(n^q_L)$ qubits to store the latent state of the evolving system. The observed register uses $\log_2(n_O)$ qubits that will be measured and reset for each time step. First, an arbitrary initial state gate is applied to the latent register. This step is analogous to setting the initial latent state for a classical HMM. For each time step, we apply a unitary gate to the whole system, then measure and reset the observed register. The outcome of this measurement determines the emitted state of the model for its respective time step. As a result of the observation, the full system projects onto the subspace where the observed qubits match the observation result. This process is mathematically equivalent to a quantum channel of dimension $n^q_L$ with $n_O$ distinct Kraus operators, see also \cite{markov2022implementation}. We show an arbitrary example circuit for three time steps in Figure \ref{fig:circ_qhmm}. For our practical implementation, we chose to use a linear entanglement layer as our initial state, followed by an EfficientSU2 gate. We present a deconstructed one time step example circuit in Figure \ref{fig:esu2}.

There are limitations on the number of time steps one can implement with currently available QPUs. The number of time steps is limited by the noise characteristics of the processor. Current quantum processors (QPUs) face significant limitations when it comes to executing deep and complex quantum circuits. These limitations stem primarily from two sources: hardware noise and short coherence times. Coherence time, particularly $T_2$ (dephasing) time, sets an upper bound on how long quantum information can be preserved in a qubit before it decays due to interactions with the surrounding environment. Once decoherence occurs, the quantum state becomes corrupted, leading to errors in computation. An additional and compounding limitation arises from imperfect gate operations. Despite advances in fabrication and control, each quantum gate still introduces a non-negligible error. When multiple gates are applied sequentially, as is necessary for deeper circuits, these errors accumulate, often exponentially, which drastically reduces the overall fidelity of the computation. Consequently, NISQ-era devices remain practically restricted to executing circuits of relatively low depth.
Moreover, real-world hardware constraints introduce further complications during circuit transpilation. In superconducting architectures, for example, limited qubit connectivity necessitates frequent qubit swaps to facilitate two-qubit operations, artificially inflating the effective circuit depth. This added overhead not only increases susceptibility to noise but also shortens the practical timescale over which a quantum algorithm can be reliably executed.
These challenges are especially pronounced in algorithms that simulate dynamical processes or require iterative applications of parameterized circuits, such as in variational quantum algorithms or quantum machine learning routines. The depth and fidelity demand of these methods frequently exceed the current capabilities of state-of-the-art QPUs, rendering certain computational tasks infeasible without significant architectural or algorithmic advances.

\smallskip

We estimate model parameters with the Nelder-Mead numerical method because it provides a direct comparison between the classical and quantum models. The Nelder-Mead algorithm is popular for quantum applications because it is a gradient-free method. We do impose some constraints, expressed as a function $\mathcal{C}(\theta),$ on the optimization to guarantee a valid transition matrix. For classical HMM models we use $\mathscr{C}$ to prevent the optimizer from generating a model with negative parameters. In the non-parametric case, we add the following constraint to ensure that the input parameters produce a stochastic matrix, i.e.\ the sum of each group of $n^c_L-1$ parameters that become a row of our transition matrix must be less than 1.

\subsection{Hardware and Markov Properties \label{subsec:Markov}}

It worth noting - as detailed in Appendix Section \ref{appsec:QC} - that the statistical ensemble of the hidden state after having observed a history of observations up to time $t,$ denoted as $\rho^L_t$ is Markovian. A noisy quantum hardware implementation raises issues about the Markov properties of the physical process used to sample from a QHMM, while the model itself remains Markovian. It should be noted that the exact definition of what makes a general quantum process Markovian is still a matter of discussion, see \cite{Rivas2012}, \cite{Smart_2022}, among others. We use the definition from \cite{PhysRevLett.120.040405} and their proposed \textit{causal break} test to verify the the simulated model is Markovian.

\smallskip

The classical Markov property is not useful when studying real quantum systems because it is predicated on the assumption that we can directly observe the complete latent state, $\mathfrak{U}_t$ or $\rho_t$ in the classical and quantum cases respectively. On quantum hardware, this would entail recreating the model emissions that produced $\rho_t$ and performing quantum state tomography which is not a practical test.\footnote{See \cite{PhysRevLett.108.070502}, \cite{10417060} for resource estimates and examples of approximate tomography procedures.} We follow the test proposed by \cite{PhysRevLett.120.040405} which involves preparing a quantum process, making a series of operations on said process, and resetting the system to a new latent state. The act of resetting the system is called a \textit{causal break}. If the probability distribution of the system after the causal break is affected by the memory of the previous observations, then the process is non-Markovian. We formally rewrite this definition of Markovianity in the notation of this paper:
\begin{definition}[Markovian Quantum Process]
	\label{daf_markovian}
	A QHMM is Markovian when the likelihood of a given sequence after resetting the latent state is only affected by said latent state, and not previous measurement outcomes.
\end{definition}

\begin{figure}
	\begin{center}
		{\includegraphics[scale=0.5]{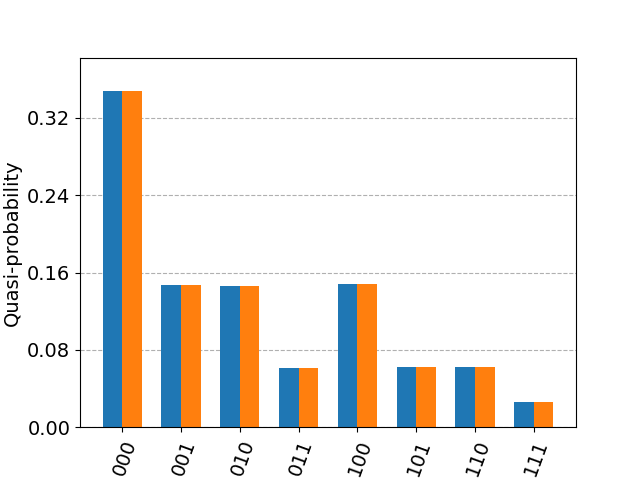}} \hskip0.1in
	\end{center}
	\caption{\label{fig:markovian} Probability distributions of the final three states from two models. The blue model represent previously emitted a sequence of $\ket{1}\ket{1}$. The orange model previously emitted $\ket{0}\ket{0}$, however we reset the latent state to be equal to $\rho_2$ of the blue model. Each model produces an identical distribution after the causal break and latent state reset, indicating that the process is entirely dependent on the current latent state, and not memory effects from prior emissions.}
\end{figure}

We perform a causal break test using two simulated QHMMs with the same circuit ansatz and rotation parameters. Figure \ref{fig:markovian} shows the probability distributions of the final three steps from two $n^q_L=4$, $n_o = 2$ models with the same model parameters. The blue histogram represents the probability of the final three emitted states of a sequence, given that the first two steps emitted $\ket{1}, \ket{1}$ and the latent state of the system evolved to $\rho_2$. The orange histogram represents the final three time steps of the second QHMM after emitting $\ket{0}, \ket{0}$, then undergoing a causal break where the latent state is reset to $\rho_2$. As expected the two distribtuions match for all possible sequences. Therefore, model itself is Markovian when perfectly simulated classically. Non-Markovian characteristics could arise from information exchanges between a noisy quantum processor an its environment. However, the theoretical results in this paper are not affected by such outside sources of non-Markovianity.  

\subsection{Complexity Analysis \label{subsec:complexity}}
The most appealing attribute of a QHMM is the quadratic reduction in the number of hidden states required to replicate a given process. The time complexity of a generic HMM is polynomial in the number of hidden states. Hence, the time complexity will be reduced by the quantum speedup. However, we also observe additional differences in runtime complexity for QHMM versus classical HMM models. Namely, we find that training a QHMM on a QPU offers an exponential speedup with respect to the number of hidden states ($n_L^c$ or $n_L^q$) used in the model. This result is counterbalanced by an exponential slowdown with respect to the time series sample length $\bf{T}.$ 
Our complexity analysis builds on results pertaining to the complexity of a Variational Quantum Algorithm (VQA). We then restructure these result in terms of QHMM parameters.\footnote{Our complexity analysis builds on the work of \cite{resch2021acceleratingvariationalquantumalgorithms} and \cite{TILLY20221}. }

VQAs operate through an iterative hybrid quantum-classical process. A quantum circuit is executed multiple times to estimate expectation values, and a classical optimizer updates the circuit parameters based on the results. The total execution time of a VQA can be expressed as $T_{total}$ = $N_{iter}  \left(T_{optimize} + N_{s} \ T_{sample}\right),$
where $N_{iter}$ represents the number of optimization iterations, which depends on the cost function landscape and optimizer efficiency. $T_{optimize}$ is the classical optimization time per iteration. $N_s$ is the number of circuit evaluations (samples) per iteration, determined by the variance in expectation values, and $T_{sample}$ denotes the quantum circuit execution time per sample. 

From a computational complexity perspective, the optimizer iterations $N_{iter}$ can scale polynomially in well-behaved cases but may become superpolynomial for highly non-convex landscapes. Further research could characterize the landscape of a QHMM likelihood function. We will find that $N_{iter}$ is comparable for the QHMM and classical HMM approaches we study. 

The required number of samples $N_s$ typically scales as $\mathcal{O}\left(1 /\epsilon^2\right)$, where $\epsilon$ is the precision in estimating expectation values. The classical optimization time $T_{optimize}$ depends on the number of parameters $P$, often giving $T_{optimize} = \mathcal{O}(P^{\alpha})$, where $\alpha$ varies based on the optimization method. 

The sampling time per quantum circuit execution, $T_{sample}$, varies depending on the QPU and classical hardware used to run the circuit. The total time to execute consists of two primary components: $T_{sample}$ = $T_{execute} + T_{process},$ where $T_{execute}$ is the quantum circuit execution time, primarily dependent on gate depth and latency. $T_{process}$ accounts for classical overhead.

The circuit execution time $\left(T_{execute}\right)$ is dependent on the quantum gate depth, scaling as:
\begin{equation*}
	T_{execute} = \mathcal{O} \left(d \ t_{gate} \right)
\end{equation*}
where $d$ is the circuit depth. For a QHMM we note that $d$ = $d_a$ $n_{time steps}$ where $d_a$ once again denotes the depth of the single time step Ansatz circuit. The time $t_{gate}$ is the average quantum gate latency, varying by hardware platform.

The processing delay $T_{process}$ consists of classical pre- and post-processing, including data aggregation and measurement readout. This often scales as $\mathcal{O}\left(1\right)$, or at most $\mathcal{O}\left(\log n \right)$ for circuits with $n$ qubits. Typically, $T_{process}\ll T_{execute}$, meaning that circuit execution time dominates the total sampling time.
By incorporating these factors, we can approximate the total complexity of the proposed algorithm as:
\begin{equation}
	\label{eq:vqa_complexity}
	T_{total} = \mathcal{O} \left(N_{iter} \ P^{\alpha} \ \frac{1}{\epsilon^2}  \ poly(n) \right)
\end{equation}
where $N_{iter}$ depends on the optimization landscape, $P$ often scales with the number of qubits $n$, $1/\epsilon^2$ accounts for the precision in estimating expectation values, and $poly(n)$ represents the circuit execution time.

To translate the general VQA complexity in equation (\ref{eq:vqa_complexity}) into terms that are relevant for both quantum and classical HMMs, we observe that the required precision $\epsilon$ needs to be less than the likelihood of the generated sequence. We can approximate that, on average, the initial parameters of the QHMM will create a completely level distribution. Thus, the minimum value of $\epsilon$ such that we measure a likelihood greater than zero scales with $n_O^{-\bf{T}}$ where $n_O$ is the number of observation bins, and $\bf{T}$ is the number of time steps in the fitting sequence. We also note that the number of qubits $n$, and parameters $P$ scale with the log of $n^q_L$ for standard Ansatz circuit structures. This yields an ultimate runtime $T_{Q}$ of the proposed QHMM fitting procedure that scales with 
\begin{equation}
	\label{eq:qhmm_complexity}
	T_{Q} = \mathcal{O} \left(N_{iter}  \ polylog(n_L^q) \ poly(n_O) \ exp(\bf{T}) \right).
\end{equation}

Classical HMMs can be estimated using an array of methods, most commonly the Viterbi or Baum-Welch algorithms. For the sake of consistency we chose the same numerical optimizer (Nelder Mead) for our QHMM and classical HMM parameter estimation. For each iteration of the optimization process, we calculate the log likelihood of the sequence with the forward algorithm, which has a time complexity of $\mathcal{O} \left( poly (n_L^{c}) \bf{T} \right)$. Repeating this process for each iteration of the optimizer results in a complexity of 
\begin{equation}
	\label{eq:class_complexity}
	T_{HMM} = \mathcal{O} \left(N_{iter} \ poly (n_L^{c}) \ \bf{T} \right),
\end{equation}
where $N_{iter}$ is comparable to that of the QHMM example. The time complexity of the more popular classical approaches is $T_{V}$ = $T_{BM}$ = $\mathcal{O} \left(poly(n_L^c) \ \bf{T} \right),$
where $T_{V}$ and $T_{BM}$ are the scaling of the Viterbi and Baum-Welch approaches respectively (see \cite{1054010} and \cite{Durbin_Eddy_Krogh_Mitchison_1998} for a thorough analysis).

Ultimately, we see that a QHMM executed on a QPU offers an exponential speedup over a classical HMM with respect to the number of latent states used in the model. This theoretical advantage is not felt in practice when the number of time steps of the generated sequence is greater than the number of hidden states. However, this exciting result does suggest a potential use case where a QHMM would be the preferred choice because of both the theoretical advantage of its non-asymptotic bounds and its runtime behavior. 

\subsection{Numerical findings \label{subsec:numerical}}

\begin{figure}
	\begin{center}
		\vskip0.1in \setcounter{subfigure}{0} 
		\vskip0.1in \subfigure[{\textsc{\scriptsize QHMM with $n_L^q$ = 2, Non-param.\ HMM with $n_L^c$ = 4}}]{\includegraphics[scale=0.35]{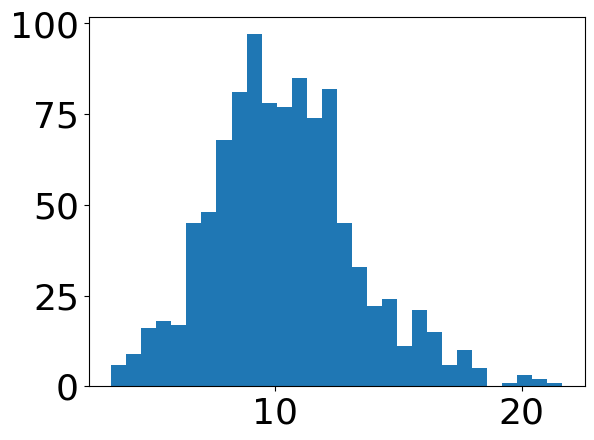}} \hskip0.1in
		\subfigure[{\textsc{\scriptsize QHMM with $n_L^q$ = 4, Non-param.\ HMM with $n_L^c$ = 16}}]{\includegraphics[scale=0.35]{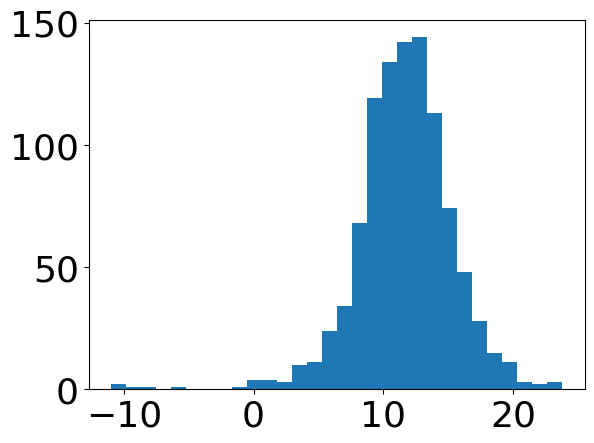}}
		\subfigure[\textsc{\scriptsize QHMM with $n_L^q$ =4, Param.\ HMM with $n_L^c$ = 16,}]{\includegraphics[scale=0.35]{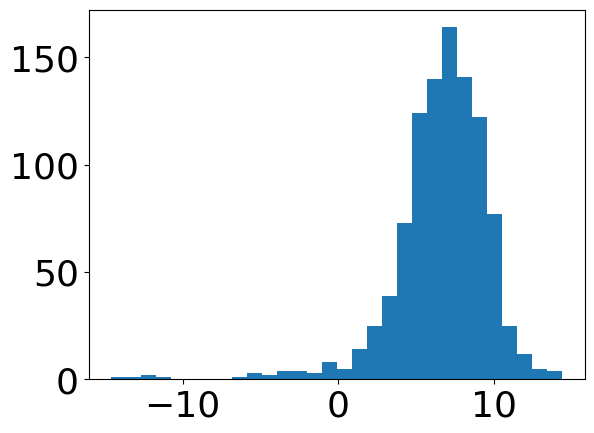}} \hskip0.1in
	\end{center}
	\caption{\label{fig:numresults} Log Likelihood Ratio (LLR) of the estimated models in each of the 1000 trials. The horizontal axis we show the LLR $\log_{10} \left(\mathscr{L}_i / \mathscr{L}_j\right)$of the trial sequence under the labeled models $i$ and $j$. The vertical axes denotes the number of trial with said LLR. 
	In panel (a) $i$ is a  QHMM with 2 latent states, $j$ is a non-parametric classical with 4 latent states, in panel (b) $i$ is a QHMM with 4 latent states and $j$ is a non-parametric classical with 16 latent states, and finally in panel (c) $i$ is a QHMM with 4 latent states, $j$ is a parametric classical one with 16 latent states.}
\end{figure}

Our numerical comparisons are based on a realistic DGP estimated from a sample of daily log returns of the S\&P 500 from 1990-2024, using the CIR model with parameters $\alpha = 2.2$, $\beta = 0.077$, and $\sigma = 1.1$ (details appear in Appendix Section \ref{appsec:implementation}).  In each trial we generate a sequence with length $\textbf{T}$ = $500,$ and estimate respectively a QHMM and a classical non-parametric or parametric HMM on samples of simulated DGPs. We repeat this process for 1000 trials and compare the resulting likelihoods. Figure \ref{fig:numresults} shows the Log Likelihood Ratio (LLR) between three pairs of models over 1000 trials. 
In panel (a) we compare a  QHMM with 2 latent states and a non-parametric classical with 4 latent states, in panel (b) a QHMM with 4 latent states and a non-parametric classical with 16 latent states, and finally in panel (c) a QHMM with 4 latent states, and a parametric classical one with 16 latent states.

\smallskip

The three panels of Figure \ref{fig:numresults} pertain to Theorem \ref{theorem:quadraticLR}, and indeed confirm the claim that for a classical HMM with $n_L^c$ = $n^2$ hidden states $\mathbb{E}^*\left[\log (\mathscr{L}(\Delta \mathscr{Y}_{t=1}^\mathbf{T}; M_{n^q_L})/	\mathscr{L}(\Delta \mathscr{Y}_{t=1}^\mathbf{T}; \tilde M_{n^c_L}))\right]$ $\geq$ 0, or equivalently: $D_{KL}^*(M_{n^q_L})$ $\leq$ $D_{KL}^*(\tilde{M}_{n^c_L})$ where the QHMM  has $n_L^q$ = $n$ hidden states.

\smallskip

Each of the panels shed a different light on the result of Theorem \ref{theorem:quadraticLR}. The first thing to note is that the statement in Theorem \ref{theorem:quadraticLR} is a large sample result, whereas  Figure \ref{fig:numresults} shows results for a modest $\textbf{T}$ = $500.$ The fact we get predictions mostly in line with the asymptotic arguments is of course encouraging. In panel (a), covering the case of a QHMM with $n_L^q$ = 2, and a non-parametric HMM with $n_L^c$ = 4, features no negative LLR outcomes. This means the QHMM is always closer in terms of KL-divergence to the true DGP. 
As the number of hidden states increases, depicted in panel (b), where $n_L^q$ = 4, and $n_L^c$ = 16, we see a small number of negative outcomes, more precisely 8 out of 1000, i.e.\ less than 1 \%. It is worth noting that the non-parametric classical HMM involves 240 parameters (to populate the transition matrix) in this case (more on this in the next section). Panel (c) is most remarkable as it compares the same type of models, except that the classical HMM is one with the same number of hidden states but only three parameters as it assumes the CIR model to characterize the transition matrix. Here again, we observe that the asymptotic result of Theorem \ref{theorem:quadraticLR} remain mostly valid in small samples, and this without the parameter proliferation of the nonparametric classical HMM. Of course, Theorem \ref{theorem:quadraticLR} pertains to KL-divergence, not the number of parameters. Therefore the result in panel (c) is not entirely surprising. Note that in panel (c) there are 31 out of 1000 cases where the classical parametric HMM is closer to the true DGP, a very small number given the fact that we have a classical model that matches the DGP.

\setcounter{equation}{0}
\section{Non-asymptotic Bounds \label{sec:nonasymp}}

The simulation results displayed in Figure \ref{fig:numresults} pertain to the large sample theory of Theorem \ref{theorem:quadraticLR}, but also suggest that there might more to explore in terms of non-asymptotic analysis.

\smallskip

We focus here on what we called the non-parametric approach, where we don't specify the drift and volatility functions of the SV diffusion but instead approximate the transition kernel with an expanding discrete state Markov chain. Since the HMMs, classical or quantum, have discretely valued hidden states they are at best approximations to the DGP. As we estimate these approximate HMMs via MLE we are obviously not dealing with inference using the true model. Panel (b) in Figure \ref{fig:numresults} highlighted the fact that the classical HMM with $n_L^c$ = 16 involves 240 parameters (to populate the transition matrix). This suggests we should consider a penalized estimation approach.

\smallskip

For the classical HMM, the parameter vector determines the transition matrix $\mathscr{A}$ and $\mathcal{X}_0,$ and any estimator will be denoted by $\hat{M}_{\mathbf{T}}(n_L^c)$ with model $M$ involving $n_L^c$ hidden states, and sample size $\mathbf{T}.$ We will be looking at a countable family of models  $\mathscr{M}_{n_L^c,\mathbf{T}}$ which may depend on the sample size and selecting one of them by choosing $\hat{\mathscr{A}}_{M,\mathbf{T}}(n_L^c)$ and $\hat{\mathcal{X}}_{0;M,\mathbf{T}}(n_L^c).$ The larger the set of models, the more distributions can be approximated, but to avoid parameter proliferation we penalize the likelihood function appearing in equation (\ref{eq:log-likeDeltay}) using a penalty function $\Lambda(n_L^c, M_{n_L^c}).$ 
To that end, we define the set $\mathscr{I}$ $=:$ $\{n^2, n \in \mathbb{N}\}.$ We will select classical HMM with $n_L^c$ $\in$ $\mathscr{I},$ such that when we take $\sqrt{n_L^c}$ we get an integer number of states for the KL-equidivergent quantum HMM. More specifically, we consider the estimator using the  likelihood function defined in (\ref{eq:log-likeDeltay}) and the associated estimator:
\begin{equation}
		\label{eq:log-likeDeltay-estim}
	(\hat{n}_L^c, \hat{\mathscr{A}}_{M,\mathbf{T}}(n_L^c), \hat{\mathcal{X}}_{0;M,\mathbf{T}}(n_L^c)) \in  \underset{n_L \in \mathscr{I},  M \in \mathscr{M}_{n_L^c,\mathbf{T}}}{\text{argmax}} \left( \frac{1}{\mathbf{T}} 	
	\mathscr{L}(\Delta \mathscr{Y}_{t=1}^\mathbf{T}; M) - \Lambda_{\mathbf{T}}(n_L^c,M) \right)
\end{equation}
yielding the non-parametric penalized maximum likelihood estimator for the classical HMM which we will denote by $\hat{M}_{\mathbf{T}}(n_L^c)$ as a shorthand for $(\hat{n}_L^c, \hat{\mathscr{A}}_{M,\mathbf{T}}(n_L^c), \hat{\mathcal{X}}_{0;M,\mathbf{T}}(n_L^c)).$ To establish non-asymptotic bounds, we rely on Theorem 6 of \cite{lehericy2021nonasymptotic}, namely:
\begin{thm}
	\label{thm:nab} Let $\mathbb{NAB}_P(\tau,\mathbf{T},n_L^c, \hat{M}_{n_L^c})$ be non-asymptotic bound for 
$D_{KL}^*(\hat{M}_{n_L^c})$ specified in Appendix equation (\ref{appeq:EJS1v2}). Let $\mathbb{NAB}_Q(\tau,\mathbf{T},\sqrt{n_L^c},\mathcal{M}^{q}_{\sqrt{n^c_L},\mathbf{T}}(\hat{M}_{n_L^c}))$ be the non-asymptotic bound for the quantum HMM equidivergent to $\hat{M}_{n_L^c}$ specified in Appendix equation (\ref{appeq:EJS1v3}).
Let Assumptions \ref{assum:compact}  through \ref{assum:constants} hold. 
Then for all $\mathbf{T} \geq n_0$, $\tau \geq 1$ and $\eta \leq 1$, with probability at least $1 - e^{-\tau} - 2 \mathbf{T}^{-2}$,
\begin{eqnarray}
	\label{eq:EJS6}
		D_{KL}^*(\hat{M}_{n_L^c})
	& \leq & \mathbb{NAB}_Q(\tau,\mathbf{T},\sqrt{n_L^c},\mathcal{M}^{q}_{\sqrt{n^c_L},\mathbf{T}}(\hat{M}_{n_L^c})) + f(\mathbf{T})\text{poly}\left((n_L^c)^2\right) \notag \\
	D_{KL}^*(\mathcal{M}^{q}_{\sqrt{n^c_L},\mathbf{T}}(\hat{M}_{n_L^c}))
	& \leq & \mathbb{NAB}_Q(\tau,\mathbf{T},\sqrt{n_L^c},\mathcal{M}^{q}_{\sqrt{n^c_L},\mathbf{T}}(\hat{M}_{n_L^c})) 
\end{eqnarray}
with explicit expressions for $f(\mathbf{T})$ and $\text{poly}\left((n_L^c)^2\right)$ appearing in Appendix equation (\ref{appeq:finalbound}).
\end{thm}
The above theorem tells us that for every classical HMM model we estimate, we have a KL-equidivergent quantum HMM with a tighter non-asymptotic bound, i.e.\ yields better estimates.\footnote{Note the observation in Remark \ref{remark:shaodow} is relevant here as well. For further discussion of concentration inequalities appearing in Theorem \ref{thm:nab} in the context of quantum statistics, see e.g.\ \cite{girotti2023concentration}.} This becomes even clearer when we express the KL divergence in terms of the filtering of the latent integrated volatility process, as shown in the next corollary.
\begin{corol}
	\label{cor:volfilter}
	For all $t$ let $\bar{V}_t^*$ be the true integrated volatility $\int_{(t-1)\Delta}^{t\Delta} V_\tau d \tau$ appearing in equation (\ref{eq:MarkovSV}), and let $V_t(\hat M)$ be the filtered integrated volatility given estimated model $\hat M.$
	Let the conditions in Theorem \ref{thm:nab} hold. Then for all $\mathbf{T} \geq n_0$, $\tau \geq 1$ and $\eta \leq 1$, with probability at least $1 - e^{-\tau} - 2 \mathbf{T}^{-2}:$
\begin{eqnarray}
	\label{eq:EJS7}
	\frac{1}{2\mathbf{T}}\sum \left(\frac{\bar{V}_t^*}{V_t(\hat{M}_{n_L^c})} - 1 \right) 
	& \leq & \mathbb{NAB}_Q(\tau,\mathbf{T},\sqrt{n_L^c},\mathcal{M}^{q}_{\sqrt{n^c_L},\mathbf{T}}(\hat{M}_{n_L^c})) + f(\mathbf{T})\text{poly}\left((n_L^c)^2\right) \notag \\
		\frac{1}{2\mathbf{T}}\sum \left(\frac{\bar{V}_t^*}{V_t(\mathcal{M}^{q}_{\sqrt{n^c_L},\mathbf{T}}(\hat{M}_{n_L^c}))} - 1 \right) 
	& \leq & \mathbb{NAB}_Q(\tau,\mathbf{T},\sqrt{n_L^c},\mathcal{M}^{q}_{\sqrt{n^c_L},\mathbf{T}}(\hat{M}_{n_L^c})) 
\end{eqnarray}	
\end{corol}
Equation (\ref{eq:EJS7})  reveals that the estimation error of the KL-equidivergent quantum HMM for filtered volatility has a tighter probability bound.

\section{Future Research \label{sec:future}}

The models studied in this paper are approximate to facilitate the computation of the likelihood function and the filtering of latent volatility. Econometricians have a long tradition of dealing with  statistical modeling when the true likelihood function is either unknown or difficult to compute. Quasi-MLE (QMLE) utilizes a surrogate likelihood function that approximates the true likelihood, allowing for consistent estimation of model parameters typically in situations where the error distribution is not accurately modeled. QMLE, studied by \cite{whi82} and \cite{gourieroux1984pseudo} among others. Although QMLE does not result in fully efficient estimates under misspecification, it often provides good asymptotic properties and consistency, making it a valuable tool for inference in complex or partially specified models.

\medskip

Maximum likelihood estimation (MLE) is the most widely used approach in quantum statistics and the quantum Cram{\'e}r-Rao (CR) bound of \cite{helstrom1967minimum} and \cite{holevo1973statistical}, and the related concept of the Quantum Fisher Information (QFI) may be regarded as the starting point of quantum estimation theory. Since then considerable progress has been made on the estimation of multi-parameter models in the context of a correctly specified likelihood function, see \cite{demkowicz2020multi} for a recent survey. The idea of a correctly specified likelihood function is justified as the empirical data in physics is typically experimental and the design is such that the model matches the data. The classical LAN asymptotic theory \cite{lecam1960locally}, and its extensions to quantum statistical models, see e.g.\ \cite{guctua2006local}, \cite{guctua2007local}, among others, both assume that the likelihood function is correctly specified. 
The important next step is the development of a theory of quantum QMLE, or q-QMLE. There are some hints about it in recent work by \cite{yano2024statistical} but there is no fully developed asymptotic theory yet. This is currently a topic of ongoing research (see \cite{ghysels_q-qmle}).

\medskip

Future research obviously also depends on the development of hardware. The practical implementation of deep quantum circuits is currently constrained by the coherence times of qubits and the architectural limitations of quantum processors. While present-day quantum hardware can support only relatively shallow circuits with acceptable fidelity, executing deeper and more expressive circuits remains a substantial challenge. Overcoming these limitations will require advances in qubit coherence, reduction of gate errors, improved connectivity, and the development of scalable error mitigation and correction techniques. Nevertheless, despite these hardware constraints, the algorithm presented here demonstrates a promising computational framework. Our results suggest that, once more capable hardware becomes available, particularly systems that can reliably support longer coherent evolutions and greater circuit depths, more time steps can be executed through the hardware. This positions our method as a forward-looking contribution, designed to take advantage of the next generation of quantum processors as they emerge. As quantum technology matures, we expect this approach to transition from simulation-based feasibility studies to impactful, hardware-executed solutions for real-world problems.

\newpage
\bibliographystyle{econometrica}
\ifx\undefined\BySame
\newcommand{\BySame}{\leavevmode\rule[.5ex]{3em}{.5pt}\ }
\fi
\ifx\undefined\textsc
\newcommand{\textsc}[1]{{\sc #1}}
\newcommand{\emph}[1]{{\em #1\/}}
\let\tmpsmall\small
\renewcommand{\small}{\tmpsmall\sc}
\fi


\clearpage \newpage
\begin{center}
	\LARGE \textbf{Appendix}
\end{center}
\onehalfspacing
\renewcommand\thepage{Appendix - \arabic{page}}
\setcounter{page}{1}
\setcounter{section}{0}
\setcounter{equation}{0}
\setcounter{table}{0}
\setcounter{figure}{0}
\setcounter{lemma}{0} 
\setcounter{proposition}{0} 
\setcounter{assumption}{0} 
\setcounter{thm}{0}
\renewcommand{\theequation}{A.\arabic{equation}}
\renewcommand\thetable{A.\arabic{table}}
\renewcommand\thefigure{A.\arabic{figure}}
\renewcommand\thesection{A.\arabic{section}}
\renewcommand{\thesection}{A.\arabic{section}}
\renewcommand{\theproposition}{A.\arabic{proposition}} 
\renewcommand{\theassumption}{A.\arabic{assumption}} 
\renewcommand{\thelemma}{A.\arabic{lemma}} 
\renewcommand{\thethm}{A.\arabic{thm}} 
{\small 

\section{Regularity Conditions and Technical Details \label{appsec:regcond}}

We detail the regularity conditions pertaining to the data generating process and the approximate hidden Markov models. We start with the former. A subsection is dedicated to each.

\subsection{Regularity Conditions - Data Generating process}

We start with an assumption which simplifies some of the derivations, but can be relaxed without loss of generality.

\begin{assumption}
	\label{assum:compact}
	Consider the sets $\{Y_\tau: \tau \in \mathcal{T} \subset \mathbb{R}^+, y_\tau \in S_Y \subset \mathbb{R}\}$ pertaining to the process  characterized by the stochastic differential equation (\ref{eq:SDE}), its discrete  time record $y_t$ $\equiv$ $Y_{t \Delta}$ for some fixed time increments $\Delta,$ yielding $\{y_t: t \in T \subset \mathbb{N}, y_t \in S_y \subset \mathbb{R}\}$ and $S_{\Delta y}$ appearing in Definition \ref{def_obs}.
	The sets $S_Y,$ $S_y,$ and $S_{\Delta y}$ are compact subsets of $\mathbb{R}.$ The processes $\bar{V}_t$ and $V_{t\Delta}$ in equation (\ref{eq:MarkovSV}) are defined on respectively $S_{\bar{V}}$ and $S_V,$ both compact subsets of $\mathbb{R}^+.$ 
\end{assumption}

The next set of regularity conditions are taken from \cite{genon2000stochastic} in their analysis of SV diffusion models.

\begin{assumption}
	\label{assum:A0}
$(B_\tau, W_\tau)_{\tau \geq 0}$ a two-dimensional standard Brownian motion in $\mathbb{R}^2,$ defined on a probability space $(\Omega, \mathcal{F}, \mathbb{P}),$ and $V_0$ defined on $\Omega$ independent of $(B_\tau, W_\tau)_{\tau \geq 0}.$ 
\end{assumption}

\begin{assumption}
	\label{assum:A1}
 Consider the interval $(l, r)$ with  $(-\infty \leq l < r \leq \infty).$ The functions $a(\cdot)$ and  $b(\cdot)$ in the SV diffusion model (\ref{eq:SDE}) are defined on $(l, r),$ and satisfy:
 \begin{itemize}
 	\item $a(x)$ $\in$ $C^1(l, r),$ $b^2(x)$ $\in$ $C^2(l, r),$ $b(x)$ $>$ 0 for all $x$ $\in$ $(l, r)$
 	\item $\exists$ $K$ $>$ 0, $\forall$ $x$ $\in$ $(l, r),$ $\vert a(x) \vert$ $\leq$ $K(1+x)$ and $b(x)^2$ $\leq$ $K(1+x^2)$ 
 \end{itemize} 
\end{assumption}

\begin{assumption}
	\label{assum:a2}
	For $\tilde{x}$ 	$\in$ $(l, r)$ define the scale and speed densities of diffusion (\ref{eq:SDE}),
	\[
	s(x) = \exp{\left(-2 \int_{\tilde{x}}^x \frac{a(u)}{b^2(u)} du \right)}, \qquad m(x) = \frac{1}{b^2(x) s(x)}.
	\]
	Then, the following hold: (a) $\int_l s(x) dx$ = $+\infty,$ (b) $\int^r  s(x) dx$ = $+\infty,$ and (c) $\int^r_l  m(x) dx$ = $M$ $<$ $+\infty.$ 
\end{assumption}

\begin{assumption}
	\label{assum:a3}
	Define the stationary density:
	\[
	\mathcal{X}_0(x) = \frac{1}{M} m(x) \mathbf{1}_{x \in (l,r)}
	\]
	The initial random variable $V_0$ has distribution $\mathcal{X}_0(dx)$ = $\mathcal{X}_0(x) dx$
\end{assumption}

Next, we list additional regularity conditions imposed by \cite{lehericy2021nonasymptotic}, namely let $\mathbb{E}^*[.]$ be the expectation under the true DGP. Likewise, $\ell^*$ is the likelihood under the true DGP. A first assumption is that the log-density rarely takes extreme values:

\begin{assumption}
	\label{assum:Astar_tail}
	There exists $\delta > 0$ such that 
	\begin{equation*}
		M_\delta := \sup_{t,k} \mathbb{E}^*[ (\log \ell^*(\Delta y_t | \Delta \mathcal{Y}_{i-k}^{i-1}))^\delta ] < \infty.
	\end{equation*}
\end{assumption}

\noindent Assumption \ref{assum:Astar_tail} pertains to the tail behavior of returns. Since we did not commit to a specific diffusion model, we use the generic assumption of \cite{lehericy2021nonasymptotic}, which he calls A$\star$tail. If we were to specify a specific diffusion, then we can refine and tailor the assumption according to the assumed DGP. See \cite{kluppelberg2010stochastic} for further discussion about tail behavior implied by specific SV models.

\medskip

Proposition 3.2 of \cite{genon2000stochastic} establishes ergodicity and $\alpha$-mixing under Assumptions \ref{assum:A0} through \ref{assum:a3}. However,
\cite{lehericy2021nonasymptotic} assumes that the DGP is $\rho$-mixing,  which is used to obtain Bernstein-like concentration inequalities. Note that $\rho$-mixing implies $\alpha$-mixing and ergodicity,

\begin{assumption}
	\label{assum:Astar_mixing}
The process	$\{\Delta y_t, t \in T\}$ is $\rho$-mixing. The $\rho$-mixing coefficient  is defined by
\begin{equation*}
	\rho_\text{mix}(T) = \rho_\text{mix}(\sigma(\Delta y_t,  \geq T), \sigma(\Delta y_t, t \leq 0)).
\end{equation*}
Then, there exist two constants $c_* > 0$ and $T_* \in \mathbb{N}$ such that
	\begin{equation*}
		\forall T \geq T_*, \quad \rho_\text{mix}(T) \leq 4 e^{- c_* n}.
	\end{equation*}
\end{assumption}

\begin{assumption}
	\label{assum:Astar_forgetting} There exist two constants $C_* > 0$ and $\rho_* \in (0,1)$ such that for all $ t \in T \subset \mathbb{N},$, for all $k,$ $k^\prime$ $\in$  $\mathbb{N},$ and for all $\Delta y_{t-(k \vee k^\prime)}^t$,
	\begin{equation*}
		| \log \ell^*(\Delta y_t | \Delta \mathcal{Y}_{t-k}^{t-1}) - \log \ell^*(\Delta y_t | \Delta \mathcal{Y}_{t-k^\prime}^{t-1}) | \leq C_* \rho_*^{k \wedge k^\prime - 1}
	\end{equation*}
\end{assumption}

\noindent Assumption~\ref{assum:Astar_forgetting} ensures that the process forgets its initial distribution exponentially fast, which is not guaranteed by $\rho$-mixing.

\subsection{Regularity Conditions - Classical HMM}

We present the regularity conditions for classical hidden Markov models used by \cite{lehericy2021nonasymptotic} to derive the non-asymptotic bounds appearing in Section \ref{sec:nonasymp}.
Recall that the classical HMM according Definition 	\ref{def_chmm} is defined by a 5-tuple: $\left(\mathcal{S}_O,\mathcal{H}_L, \mathscr{A}, \mathscr{E}, \mathcal{X}_0\right),$
where $\mathcal{X}_0$ is a
stochastic vector defining the initial latent process states and $\mathscr{E}$ the emission matrix.  A shorthand notation for models will be $M_{n_L^c}$ simply emphasizing the number of hidden states. First, two assumptions that apply to both classical and quantum HMM.

\begin{assumption}
	\label{assum:knownhidden} The number of hidden states, $n_L^c$ for HMM and $n_L^q$  for QHMM, is assumed known or preset.
\end{assumption}

\begin{assumption}
	\label{assum:orderrank} The order of a HMM, namely the dimensionality of the hidden state space, is equal to the rank of the HMM. 
\end{assumption}

\begin{assumption}
	\label{assum:naturalorder}
	The partition set $\mathcal{B}_O$ appearing in Definition \ref{def_obs} has a natural order, starting with left-tail observation in $b_1$ and right-tail ones in $b_{n_o}.$ This yields a natural order for realizations $\mathfrak{y}_{t}$ = $\iota_i$ where $\iota_i$ is the $i^{th}$ coordinate vector of $\mathbb{R}^{n_o}.$ We will use the labels of bins and refer to these as $\iota_1$ through  $\iota_{n_o}.$
\end{assumption}

In the remainder of this section we  state the assumptions in terms of the $\Delta \mathscr{Y}_{t=1}^\mathbf{T}$ data.
The first assumption imposes bounds on the transition matrix and by implication implies that the difference between the log-likelihood appearing in equation	(\ref{eq:log-likeDeltay}) normalized by the sample size, namely
$\mathscr{L}(\Delta \mathscr{Y}_{t=1}^\mathbf{T}; M_{n_L^c})/\mathbf{T}$ and its limit converges to zero with rate $1/\mathbf{T}$ in supremum norm.
\begin{assumption}
	\label{assum:Aergodic} There exists $C_{\mathbf{A}} \geq 1$ such that for all $(n_L^c, \mathcal{X}_0, \mathbf{A}, \mathscr{E}) \in M_{n_L^c}$,
	\begin{eqnarray*}
		\forall x, x' \in [n_L^c], & & (C_{\mathbf{A}} \log \mathbf{T})^{-1} \leq n_L^c \mathscr{A}(x,x') \leq C_{\mathbf{A}} \log \mathbf{T} \\
		& & \\
		\forall x \in [n_L^c], & &  (C_{\mathbf{A}} \log \mathbf{T})^{-1} \leq n_L^c \mathcal{X}_0(x) \leq C_{\mathbf{A}} \log \mathbf{T}.
	\end{eqnarray*}
\end{assumption}
\noindent The following assumption ensures that the log-likelihood rarely takes extreme values. In our case, since all emission probabilities are Gaussian, this assumption is satisfied by the approximate hidden Markov models.
\begin{assumption}
	\label{assum:Atail}
	There exists $C_{\ell^*} \geq 1$ such that
	\begin{equation*}
		\forall u \geq 1, \quad \mathscr{P}^* \left[ \sup_{\mathscr{E} \in M_{n_L^c}^{(\mathscr{E})}} \left| b_\mathscr{E}(\Delta y) \right| \geq C_{\ell^*} (\log \mathbf{T}) u \right] \leq e^{-u}.
	\end{equation*}
where 
$\mathscr{P}^*$ is the probability under the true DGP and for all $\mathscr{E} \in M_{n_L^c}^{(\mathscr{E})}$ and $\Delta y \in S_{\Delta y}$, let
$b_\mathscr{E}(\Delta y)$ = $\log \left((n_L^c)^{-1} \sum_{x} \mathscr{E}_x(\Delta y) \right).$
\end{assumption}

\begin{remark}
	\label{appremark:entropy}
	The framework in \cite{lehericy2021nonasymptotic} involves parametric emission densities. To ensures that the emission densities behave like finite-dimensional parametric models, \cite{lehericy2021nonasymptotic} imposes an assumption called {\it Aentropy}. Since in our case the emissions do not involve any additional parameters and the dimensionality of the model coincides with the number of hidden states, we simply assume there is a constant $C_{aux} \geq 1$ instead of a model-dependent function subject to bracketing entropy restrictions. It also implies we do not need to impose {\it Agrowth} assumption on the function.
\end{remark}

\begin{assumption}
	\label{assum:constants}
	The constants $A$ and $C_{\Lambda}$ depending only on $C_{\ell^*}$, mixing parameters $T_*$ and $c_*$ and a constant $\mathbf{T}_0$ depending only on $C_{\ell^*}$, $T_*$, $C_*$, $\rho_*$, $\delta$ and $M_\delta$ which apply to the classical HMM non-asymptotic bound also apply to the KL-equidivergent quantum HMM.
\end{assumption}

In light of Theorems \ref{lemma:quadraticdim} and \ref{theorem:quadraticdimred} the above assumption is reasonable to make and allows us to make direct comparisons of the bounds in Theorem \ref{thm:nab}.

\subsection{Regularity Conditions - Quantum HMM}

According to Definition 	\ref{def_unitary_qhmm}
a unitary Quantum HMM is defined by a 6-tuple: $\left(\mathcal{S}_O,\hilbert_L, \hilbert_O, U, \mathcal{M}, R_0\right),$
where: (a) $\mathcal{S}_O$ output states, (b) $\hilbert_L$  the tensor product space associated with hidden quantum states, (c) $\hilbert_O$  the tensor product space associated with the observable states, (d) $U$ is a unitary operator defined on the bipartite Hilbert space $\hilbert_L \otimes \hilbert_O,$ (e) $\mathcal{M}$ is a bijective map $\mathcal{P}_i^{O} \rightarrow \mathcal{S}_O$, and finally (f) $ R_0 = \rho_0^L \otimes \ket{u_0}\bra{u_0}$ is an initial emission pure state.

\medskip

We rely on two important results, due to respectively \cite{monras2011hidden} and \cite{markov2022implementation}, which we state without proof, namely: 
\begin{thm}[\cite{monras2011hidden}, Theorem section 4.3]
	\label{lemma:C-QHMM}
	For every classical HMM of order $n_L^c$ there exists a QHMM of the same order which generates the same
	stochastic process.
\end{thm}
\noindent and the second result:
\begin{thm}[\cite{markov2022implementation}, Theorem 3]
	\label{lemma:quadraticdim}
	Let $M_{n_L^q}$ be a QHMM with hidden state dimension $n_L^q$ and let $H^Q$ be its associated Hankel matrix, then rank$(H^Q) \leq (n_L^q)^2.$ 
\end{thm}

\medskip

One implication is that if we have a classical HMM of rank $n_L^c,$ as defined in Definition \ref{def:rankHMM}, we can find a QHMM with $n_L^q$ = $\sqrt{n_L^c}$ of hidden states with the same stochastic properties. One might call this a quadratic dimensionality reduction. By implication, the corresponding quantum HMM inherits the regularity conditions of the classical HMM.

\section{Lemmas \label{appsec:lemmas}}

\begin{lemma}[\cite{genon2000stochastic}]
	\label{lemma:GJL}
	Under Assumptions \ref{assum:A0} through \ref{assum:a3}, the hidden discrete time Markov process $\mathfrak{U}_t$  := $(\bar{V}_t, V_{t \Delta})$ defined in equation (\ref{eq:MarkovSV}) has a time homogeneous transition kernel dependent only on $V_{t \Delta}.$ Moreover, the process $\Delta y_t$ is strictly stationary and ergodic.
\end{lemma}

\bigskip

\begin{lemma}
	\label{lemma:approxbins}
	Let Assumptions \ref{assum:A0} through \ref{assum:a3} hold. Then, for any given $\eta$ $>$ 0, arbitrary small, there is $n_O$ such that:
	$D_{KL}(\ell^*_{\mathbf{T}} \parallel \ell_{\mathbf{T}}^{\mathfrak{y}}(M_{n^c_L}))$ - $D_{KL}(\ell^*_{\mathbf{T}} \parallel \ell_{\mathbf{T}}^{\Delta y}(\tilde{M}_{n^c_L}))$ = 	$D_{KL}(\ell_{\mathbf{T}}^{\Delta y}(\tilde{M}_{n^c_L}) \parallel \ell_{\mathbf{T}}^{\mathfrak{y}}(M_{n^c_L}))$ 
	= $\eta.$
\end{lemma}

\medskip

\begin{lemma}
	\label{lemma:zerodivergence}
	Let Assumption \ref{assum:LLN} hold.
	Consider two hidden Markov models, $M_1$ and $M_2,$ which are KL-equidivergent as defined in Definition 
	\ref{def:KL-equiv}, and therefore $D_{KL}^*(M_1)$ = $D_{KL}^*(M_2).$  Then as $\mathbf{T}$ $\rightarrow$ $\infty:$ 
	\begin{equation}
		\label{appeq:lemma2}
		D_{KL}^*(\mathscr{L}_{\mathbf{T}}^{\mathfrak{y}}(M_1)\parallel \mathscr{L}_{\mathbf{T}}^{\mathfrak{y}}(M_2)) :=  \mathbb{E}^*\left[\log \frac{\mathscr{L}( \mathfrak{Y}_{t=1}^\mathbf{T}, M_1)}{\mathscr{L}( \mathfrak{Y}_{t=1}^\mathbf{T}, M_2)}\right] = 0.
	\end{equation}
\end{lemma}



\section{Proofs \label{appsec:proofs}}

\subsection{Proof of Lemma \ref{lemma:GJL}}

We need to show first that under Assumptions \ref{assum:A0} - \ref{assum:a3}, the hidden discrete time Markov process $\mathfrak{U}_t$  = $(\bar{V}_t, V_{t \Delta})$ defined in equation (\ref{eq:MarkovSV}) has a time homogeneous transition density dependent only on $V_{t \Delta}.$ We use the arguments in the proof of Theorem 3.1 of \cite{genon2000stochastic} to obtain the result. 

\medskip

Let $\varphi$: $(l,r)^2$ $\rightarrow$ $\mathbb{R}$ be a bounded Borel function. The stochastic differential equation (\ref{eq:SDE}) for $V_\tau,$ and using the definition of $\bar{V},$ we have:
$\mathbb{E}\left[\varphi(\bar{V}_t, V_{t \Delta} \vert \sigma(V_s, s \leq t-1))\right]$ = $\mathbb{E}\left[\varphi(\bar{V}_1, V_\Delta \vert V_0)\right].$
This implies that the transition probability of the Markov chain $(\mathfrak{U}_t, t \geq 1)$ only depends on $V.$ 
Finally, the strict stationarity and ergodicity of $\Delta y_t$ follows from Proposition 3.1 in GJL.
\hfill $\Box$

\subsection{Proofs of Lemmas  \ref{lemma:zerodivergence} and \ref{lemma:approxbins}}

We start with the proof of Lemma \ref{lemma:approxbins}.
First we want to establish the existence of $\ell^*.$ Using Theorems 2.3 and 3.1 of \cite{genon2000stochastic}, we know that under Assumptions \ref{assum:A0} through \ref{assum:a3} the process is $\{\Delta y_t\}$ is stationary and ergodic. Ergodicity implies that $	\mathscr{L}(\Delta \mathscr{Y}_{t=1}^\mathbf{T}; M_{n_L^c})/\mathbf{T}$ $\rightarrow$ $\ell$ almost surely as $\mathbf{T}$ $\rightarrow$ $\infty$ (see \cite{barron1985strong} and Lemma 3 in \cite{lehericy2021nonasymptotic}).  Recall from equations (\ref{eq:KL1}) and (\ref{eq:KL2}) that:
\begin{equation}
	\label{appeq:KL1}
	D_{KL}(\ell^*_{\mathbf{T}} \parallel \ell_{\mathbf{T}}^{\mathfrak{y}}(M_{n^c_L})) = \mathbb{E}^*\left[\log \frac{\ell^*(\Delta \mathscr{Y}_{t=1}^\mathbf{T})}{\ell(\mathfrak{Y}_{t=1}^\mathbf{T}, M_{n^c_L})}\right]
\end{equation}
and:
\begin{equation}
	\label{appeq:KL2}
	D_{KL}(\ell^*_{\mathbf{T}} \parallel \ell_{\mathbf{T}}^{\Delta y}(\tilde{M}_{n^c_L})) = \mathbb{E}^*\left[\log \frac{\ell^*( \Delta \mathscr{Y}_{t=1}^\mathbf{T})}{	\ell( \Delta \mathscr{Y}_{t=1}^\mathbf{T}, \tilde{M}_{n^c_L})}\right]
\end{equation}
where $\mathbb{E}^*[.]$ is the expectation under the true DGP. From the above equation we have:
\begin{eqnarray}
	\label{appeq:KL3}
	D_{KL}(\ell^*_{\mathbf{T}} \parallel \ell_{\mathbf{T}}^{\mathfrak{y}}(M_{n^c_L})) & = & \mathbb{E}^*\left[\log \frac{\ell^*( \Delta \mathscr{Y}_{t=1}^\mathbf{T})}{	\ell( \Delta \mathscr{Y}_{t=1}^\mathbf{T}, \tilde{M}_{n^c_L})} \frac{	\ell( \Delta \mathscr{Y}_{t=1}^\mathbf{T}, \tilde{M}_{n^c_L})}{\ell( \mathfrak{Y}_{t=1}^\mathbf{T}, M_{n^c_L})}\right]  \\
	& = &  D_{KL}(\ell^*_{\mathbf{T}} \parallel \ell_{\mathbf{T}}^{\Delta y}(\tilde{M}_{n^c_L})) + \mathbb{E}^*\left[\log \frac{	\ell( \Delta \mathscr{Y}_{t=1}^\mathbf{T}, \tilde{M}_{n^c_L})}{\ell( \mathfrak{Y}_{t=1}^\mathbf{T}, M_{n^c_L})}\right] \notag
\end{eqnarray}
where the second term on the RHS of the second equation is the expected, under the true DGP, of the log likelihood ratio of the two discrete time filtrations:   $\{\mathcal{F}_t^{\Delta y}, t \in T\}$ and $\{\mathcal{F}_t^{\mathfrak{y}}, t \in T\}.$  

\medskip

Focusing on the second term, from equation (\ref{eq:dirac}) we know that:
\begin{equation}
	\label{appeq:dirac}
	\mathcal{P}_{\mathfrak{y}}(\mathfrak{y}_{t+1} = \iota_i \vert \mathfrak{U}_t) = \int_{S_{\Delta y}} \mathcal{P}(x \vert \mathfrak{U}_t) \delta_{b_i}(x) d x \qquad \forall \iota_i \in \mathcal{S}_O,
\end{equation}
with $\delta_{b_i}$ the Dirac delta function associated with bin $b_i$ and $\mathcal{P}(x\vert \mathfrak{U}_t)$ pertains to $\mathcal{N}(0,\bar{V}_t ).$ Recall from Definition \ref{def_obs} that the compact set $S_{\Delta y}$ is partitioned  $\mathcal{B}_O$ := $\{b_1, \ldots, b_{n_O}\}$ such that $\bigcup_i b_i$ = $S_{\Delta y}$ and $b_i$ $\cap$ $b_j$ = $\emptyset$ for $i$ $\neq$ $j$ with $n_O$ the number of bins, i.e.\ $n_O$ := $\vert \mathcal{B}_O \vert.$ The maximal length of the bins is denoted by $\ell_{max}^b.$  As the intervals shrink, we have that: $\mathcal{P}_{\mathfrak{y}}$ $\approx$ $\mathcal{P}$ in equation (\ref{appeq:dirac}) and therefore 
$\ell( \mathfrak{Y}_{t=1}^\mathbf{T}, M_{n^c_L})\ell_{max}^b$ $\approx$ 	$\ell(\Delta \mathscr{Y}_{t=1}^\mathbf{T}, \tilde{M}_{n^c_L}).$ Hence,  $\mathbb{E}^*\left[\log \ell( \Delta \mathscr{Y}_{t=1}^\mathbf{T}, \tilde{M}_{n^c_L})/\ell( \mathfrak{Y}_{t=1}^\mathbf{T}, M_{n^c_L})\right]$  becomes arbitrary small, since we take logs of values close to one. Since the bins are non-overlapping, the approximation is determined by $n_O$ and $\ell_{max}^b.$ We can simply construct bins of equal size, such that the approximation is only driven by $n_O$ = $\vert S_{\Delta y} \vert/\ell_{max}^b.$ As a result:
\begin{equation*}
	D_{KL}(\ell^*_{\mathbf{T}} \parallel \ell_{\mathbf{T}}^{\mathfrak{y}}(M_{n^c_L}))  =   D_{KL}(\ell^*_{\mathbf{T}} \parallel \ell_{\mathbf{T}}^{\Delta y}(\tilde{M}_{n^c_L})) + \eta
\end{equation*}
with $\eta$ determined by $n_O.$ Therefore, $D_{KL}(\ell_{\mathbf{T}}^{\Delta y}(\tilde{M}_{n^c_L}) \parallel \ell_{\mathbf{T}}^{\mathfrak{y}}(M_{n^c_L}))$ 
= $\eta,$ arbitrary small. \hfill $\Box$

\medskip

We conclude with the proof of Lemma \ref{lemma:zerodivergence}.  The result follows from  what is sometimes referred to as the Pythagorean identity, see \cite{csiszar1975divergence}. Namely, $	D_{KL}^*(\mathscr{L}_{\mathbf{T}}^{\mathfrak{y}}(M_1)\parallel \mathscr{L}_{\mathbf{T}}^{\mathfrak{y}}(M_2)) $ =  $D_{KL}^*(M_1)$ - $D_{KL}^*(M_2)$ = 0. \hfill $\Box$

\smallskip



\subsection{Proof of Theorem \ref{theorem:quadraticdimred}}

We start from a classical HMM with order (i.e.\ number of hidden states) equal to $n_L^c$ = $n^2$ for $n$ integer. Using Assumptions \ref{assum:knownhidden} through 
\ref{assum:orderrank}, we know
from \cite{anderson1999realization} and \cite{huang} we know that there is a minimal classical HMM with rank, defined in Definition \ref{def:rankHMM}, equal to $n_L^c$ and it is unique according to \cite{vanluyten2008equivalence}.  We denote the associated Hankel matrix, appearing in equation (\ref{eq:hankel}), as $\mathcal{H}^c.$ Theorem \ref{lemma:quadraticdim} implies that there is a corresponding QHMM with hidden state dimension (i.e.\ order) $n_L^q$ and associated Hankel matrix, denoted by $\mathcal{H}^q$ of  rank$(\mathcal{H}^q) \leq (n_L^q)^2.$ 

\medskip

The generalized Hankel matrix appearing in equation (\ref{eq:hankel}) is bi-infinite. Let us consider for any $d$ integer, the finite dimensional $d \times d$ Hankel matrices and their canonical representation:
\begin{eqnarray}
	\label{appeq:canonicalHankel}
	\mathcal{H}^c(d) & = & \mathcal{C}^c(d) \left(\begin{array}{cc} I_{n_L^c} & 0 \\ 0 & 0 \end{array}\right) \mathcal{D}^c(d)\\
	\mathcal{H}^q(d) & = & \mathcal{C}^q(d) \left(\begin{array}{cc} I_{(n_L^q)^2} & 0 \\ 0 & 0 \end{array} \right)\mathcal{D}^q(d) \qquad (n_L^q)^2 = n_L^c \notag
\end{eqnarray} 
where $\mathcal{C}^c(d),$ $\mathcal{D}^c(d),$ $\mathcal{C}^q(d)$ and $\mathcal{D}^q(d)$ are $d \times d$ invertible matrices.

\medskip

Recall that we are interested in $\mathcal{M}^{q}_{\sqrt{n^c_L},\mathbf{T}}(M),$ i.e.\ the set of quantum HMM models, with $\sqrt{n^c_L}$ hidden states, and the KL divergence with respect to the classical HMM model $M$ $\in$ $\mathcal{M}_{n^c_L,\mathbf{T}}.$
The KL divergence between hidden Markov models can be written in terms of their Hankel matrix representations. We first look at a generic case involving HMM$^1$ and HMM$^2$ with Hankel matrices $\mathcal{H}^1$ and $\mathcal{H}^2,$ respectively. Note that in the generic case $\mathcal{H}^1$ may be purely data-driven, and therefore not correspond to a specific model, such that the KL divergence is between data and a model, or between two models. Then the KL divergence can be written as (assuming again finite dimensions $d$):
\begin{equation}
	\label{appeq:KLmatrixgeneric}
	D_{KL}(\mathcal{H}^1 \parallel \mathcal{H}^2) = \sum_{i=1}^d \sum_{j=1}^d \left( \mathcal{H}^1_{ij} \log \frac{\mathcal{H}^1_{ij}}{\mathcal{H}^2_{ij}}- \mathcal{H}^1_{ij} + \mathcal{H}^2_{ij}  \right),
\end{equation}
see \cite{vanluyten2008equivalence}, \cite{finesso2010approximation}, among others, for further details. Of particular interest is the fact that $D_{KL}(\mathcal{H}^1 \parallel \mathcal{H}^2)$ = 0 if and only if $\mathcal{H}^1$ = $\mathcal{H}^2,$ see \cite{finesso2010approximation} Definition 5.3.

\medskip

Continuing with the generic case, and let Assumption \ref{assum:LLN} hold, we are interested in $D_{KL}^*(\mathcal{H}^1)$ versus $D_{KL}^*(\mathcal{H}^2),$ and we need to show they are equal, i.e.\ the classical HMM and reduced order quantum HMM are KL-equidivergent. From Definition \ref{def:KL-equiv} we know from Lemma \ref{lemma:zerodivergence} that the Pythagorean identity implies it is sufficient to show that $D_{KL}(\mathcal{H}^1 \parallel \mathcal{H}^2)$ = 0. Let $\mathcal{H}^c(d)$ in equation (\ref{appeq:canonicalHankel}) be the Hankel matrix associated with  classical HMM model $M$ $\in$ $\mathcal{M}_{n^c_L,\mathbf{T}}.$ 
Likewise, $\mathcal{H}^q(d)$ be the Hankel matrix associated with quantum HMM model corresponding to $M$ in $\mathcal{M}_{n^c_L,\mathbf{T}}.$ Then, using the canonical decomposition of the respective Hankel matrices we have that:
\begin{equation}
	D_{KL}( (\mathcal{C}^c(d))^{-1} \mathcal{H}^c(d) (\mathcal{D}^c(d))^{-1}\parallel (\mathcal{C}^q(d))^{-1} \mathcal{H}^q(d) (\mathcal{D}^q(d))^{-1}) = 0
\end{equation}
for any $d,$ and we use the convention that $0 \log(0/0)$ = 0. This means that a (computable and invertible) transformation of the Hankel matrix of the classical HMM is KL-equidivergent to a (computable and invertible) transformation of the corresponding reduced dimensionality quantum HMM. \hfill $\Box$

\subsection{Proof of Theorem \ref{theorem:quadraticLR}}

We apply again the Pythagorean identity, see \cite{csiszar1975divergence}. In the general case we have:
\[
D_{KL}^*(\tilde{M}_{n^c_L}) = D_{KL}^*(M_{n^q_L}) + \mathbb{E}^*\left[\log (\mathscr{L}(\Delta \mathscr{Y}_{t=1}^\mathbf{T}; M_{n^q_L})/	\mathscr{L}(\Delta \mathscr{Y}_{t=1}^\mathbf{T}; \tilde M_{n^c_L}))\right].
\]
For models in the set $\mathcal{M}^{q}_{\sqrt{n^c_L}}(M),$ i.e.\ the set of quantum HMM models, with $\sqrt{n^c_L}$ hidden states, KL-equidivergent to the classical HMM model $M$ $\in$ $\mathcal{M}_{n^c_L},$ we have:
$D_{KL}^*(\tilde{M}_{n^c_L})$ = $D_{KL}^*(M_{n^q_L})$ and therefore the expectation of the log likelihood ratio under the true DGP equals zero. However, $\mathcal{M}^{q}_{\sqrt{n^c_L}}(M)$ $\subset$ $\mathscr{M}_{\sqrt{n^c_L}},$ and therefore  $D_{KL}^*(M_{n^q_L})$ $\leq$ $D_{KL}^*(\tilde{M}_{n^c_L})$ in the larger set. Therefore $\mathbb{E}^*\left[\log (\mathscr{L}(\Delta \mathscr{Y}_{t=1}^\mathbf{T}; M_{n^q_L})/	\mathscr{L}(\Delta \mathscr{Y}_{t=1}^\mathbf{T}; \tilde M_{n^c_L}))\right]$ $\geq$ 0. \hfill $\Box$

\subsection{Proof of Theorem \ref{thm:nab}}

We start with Theorem 6 from \cite{lehericy2021nonasymptotic}, assuming Assumptions \ref{assum:compact}  through \ref{assum:Atail} hold.  Unlike \cite{lehericy2021nonasymptotic}, in our application the emissions $\mathscr{E}$ do not involve any parameters and are only driven by the coarseness of the  bins selected in $\mathcal{B}_O.$ Our analysis will be based on samples $\Delta \mathscr{Y}_{t=1}^\mathbf{T},$ relying on Lemma \ref{lemma:approxbins} to bypass the coarse sampling. Under regularity Assumptions \ref{assum:compact} through \ref{assum:Aergodic}, let $(w_M)_{M \in \mathscr{M}_{n_L}}$ be a nonnegative sequence such that $\sum_{M \in \mathscr{M}_{n_L}} e^{-w_M} \leq e-1$. Recall that we are looking at a countable family of parametric sets of models  $\mathscr{M}_{n_L^c,\mathbf{T}}$ which may depend on the sample size and selecting one of them by choosing $\hat{n}_L^c,$ $\hat{\mathscr{A}}_{M,\mathbf{T}}(n_L^c)$ and $\hat{\mathcal{X}}_{0;M,\mathbf{T}}(n_L^c).$ The larger the set of models, the more distributions can be approximated, but to avoid parameter proliferation we penalize the likelihood. To that end, we define the set $\mathscr{I}$ $=:$ $\{n^2, n \in \mathbb{N}\}.$ We will select classical HMM with $n_L^c$ $\in$ $\mathscr{I},$ such that when we take $\sqrt{n_L^c}$ we get an integer number of states for the KL-equidivergent quantum HMM. More specifically, we consider the estimator using the estimator appearing in equation (\ref{eq:log-likeDeltay-estim}) repeated here for convenience:
\begin{equation}
	\label{appeq:log-likeDeltay-estim}
	(\hat{n}_L^c, \hat{\mathscr{A}}_{M,\mathbf{T}}(n_L^c), \hat{\mathcal{X}}_{0;M,\mathbf{T}}(n_L^c)) \in  \underset{n_L \in \mathscr{I},  M \in \mathscr{M}_{n_L^c,\mathbf{T}}}{\text{argmax}} \left( \frac{1}{\mathbf{T}} 	
	\mathscr{L}(\Delta \mathscr{Y}_{t=1}^\mathbf{T}; M) - \Lambda_{\mathbf{T}}(n_L^c,M) \right)
\end{equation}
yielding the non-parametric penalized maximum likelihood estimator for the classical HMM which we will denote by $\hat{M}_{\mathbf{T}}(n_L^c)$ as a shorthand for $(\hat{n}_L^c, \hat{\mathscr{A}}_{M,\mathbf{T}}(n_L^c), \hat{\mathcal{X}}_{0;M,\mathbf{T}}(n_L^c)).$ Then there exist constants $A$ and $C_{\Lambda}$ depending only on $C_{\ell^*}$, mixing parameters $T_*$ and $c_*$ and a constant $\mathbf{T}_0$ depending only on $C_{\ell^*}$, $T_*$, $C_*$, $\rho_*$, $\delta$ and $M_\delta$ such that for all $\mathbf{T} \geq n_0$, $\tau \geq 1$ and $\eta \leq 1$, with probability at least $1 - e^{-\tau} - 2 \mathbf{T}^{-2}$,
	\begin{equation}
		\label{appeq:EJS1}
		D_{KL}^*(\hat{M}_{n_L^c})
		\leq (1+\eta) \underset{M \in \mathcal{M}_{n_L^c,\mathbf{T}}}{\underset{n_L^c \in \mathcal{I}, n_L^c \leq \mathbf{T}}{\inf}} \Bigg\{ \inf_{M \in \mathcal{M}_{n_L^c,\mathbf{T}}} D_{KL}^*(M) 
		+  2\Lambda_\mathbf{T}(n_L^c,M) \Bigg\}
		+ \frac{A}{\eta} \tau \frac{(\log \mathbf{T})^{10}}{\mathbf{T}}
	\end{equation}
	for
	\begin{equation}
		\label{appeq:ESJ2}
		\Lambda_\mathbf{T}(n_L^c,M) \geq
		\frac{C_{\Lambda}}{\eta} \frac{(\log \mathbf{T})^{10}}{\mathbf{T}} \Bigg\{ w_M
		+ (\log \mathbf{T})^{4} ( m_M n_L^c + (n_L^c)^2 - 1) 
 \left( (\log \mathbf{T})^3 \log \log \mathbf{T} + \log C_{aux} \right) \Bigg\}.
	\end{equation}
where $ C_{aux}$ is defined in Remark \ref{appremark:entropy} and $m_M$ is number of parameters in model $M.$ 

\medskip

By definition: $ \inf_{M \in \mathcal{M}_{n_L^c,\mathbf{T}}} D_{KL}^*(M)$ = $\inf_{M \in \mathcal{M}_{n_L^c,\mathbf{T}}} D_{KL}^*(\mathcal{M}^{q}_{\sqrt{n^c_L},\mathbf{T}}(M)).$
Assuming that the inequality in equation (\ref{appeq:ESJ2}) holds as an equality then we have that:
	\begin{equation}
	\label{appeq:ESJ3}
	\Lambda_\mathbf{T}(n_L^c,M) =
	\frac{C_{\Lambda}}{\eta} \frac{(\log \mathbf{T})^{10}}{\mathbf{T}} \Bigg\{ w_M
	+ (\log \mathbf{T})^{4} ( m_M n_L^c + (n_L^c)^2 - 1) 
	\left( (\log \mathbf{T})^3 \log \log \mathbf{T} + \log C_{aux} \right) \Bigg\}
\end{equation}
and we define:
	\begin{multline*}
	\Lambda_\mathbf{T}(\sqrt{n_L^c},\mathcal{M}^{q}_{\sqrt{n^c_L},\mathbf{T}}(M)) = \\
	\frac{C_{\Lambda}}{\eta} \frac{(\log \mathbf{T})^{10}}{\mathbf{T}} \Bigg\{ w_M
	+ (\log \mathbf{T})^{4} (\tilde{m}_M \sqrt{n_L^c} + n_L^c - 1) 
	\left( (\log \mathbf{T})^3 \log \log \mathbf{T} + \log C_{aux} \right) \Bigg\},
\end{multline*}
with $\tilde{m}_M$ the number of parameters in the KL-equidivergent quantum HMM. Therefore,
	\begin{equation}
	\label{appeq:ESJ5}
	\Lambda_\mathbf{T}(n_L^c,M) =	\Lambda_\mathbf{T}(\sqrt{n_L^c},\mathcal{M}^{q}_{\sqrt{n^c_L},\mathbf{T}}(M)) +
	\frac{C_{\Lambda}}{\eta} \frac{(\log \mathbf{T})^{17}\log \log \mathbf{T}}{\mathbf{T}}  ( (m_M - 1) n_L^c + (n_L^c)^2 -  \tilde{m}_M \sqrt{n_L^c} ) .
\end{equation}
We will write equation (\ref{appeq:EJS1}) as follows:
\begin{eqnarray}
	\label{appeq:EJS1v2}
	D_{KL}^*(\hat{M}_{n_L^c})
	& \leq & \mathbb{NAB}_P(\tau,\mathbf{T},n_L^c, \hat{M}_{n_L^c})  \\
	\mathbb{NAB}_P(\tau,\mathbf{T},n_L^c, \hat{M}_{n_L^c}) 	& =: &  (1+\eta) \underset{M \in \mathcal{M}_{n_L^c,\mathbf{T}}}{\underset{n_L^c \in \mathcal{I}, n_L^c \leq \mathbf{T}}{\inf}} \Bigg\{ \inf_{M \in \mathcal{M}_{n_L^c,\mathbf{T}}} D_{KL}^*(M) 
	+  2\Lambda_\mathbf{T}(n_L^c,M) \Bigg\} 
	+ \frac{A}{\eta} \tau \frac{(\log \mathbf{T})^{10}}{\mathbf{T}} \notag
\end{eqnarray}
where we assume that equation (\ref{appeq:ESJ2}) holds as an equality and therefore determining $\Lambda_\mathbf{T}(n_L^c,M).$ Moreover,
let $\mathcal{M}^{q}_{\sqrt{n^c_L},\mathbf{T}}(M)$ be the set of quantum HMM models with $\sqrt{n^c_L}$ hidden states KL-equidivergent to model classical HMM model $M$ $\in$ $\mathcal{M}_{n^c_L,\mathbf{T}},$ as stated in Theorem \ref{theorem:quadraticdimred}. By analogy, under Assumption \ref{assum:constants}, for the quantum HMM we have:
\begin{eqnarray}
	\label{appeq:EJS1v3}
	D_{KL}^*(\mathcal{M}^{q}_{\sqrt{n^c_L},\mathbf{T}}(\hat{M}_{n_L^c}))
	& \leq & \mathbb{NAB}_Q(\tau,\mathbf{T},\sqrt{n_L^c},\mathcal{M}^{q}_{\sqrt{n^c_L},\mathbf{T}}(\hat{M}_{n_L^c})) \\
	\mathbb{NAB}_Q(\tau,\mathbf{T},\sqrt{n_L^c}, \mathcal{M}^{q}_{\sqrt{n^c_L},\mathbf{T}}(\hat{M}_{n_L^c})) 	& =: &  (1+\eta) \underset{M \in \mathcal{M}_{n_L^c,\mathbf{T}}}{\underset{n_L^c \in \mathcal{I}, n_L^c \leq \mathbf{T}}{\inf}} \Bigg\{ \inf_{M \in \mathcal{M}_{n_L^c,\mathbf{T}}} D_{KL}^*(\mathcal{M}^{q}_{\sqrt{n^c_L},\mathbf{T}}(\hat{M}_{n_L^c})) \notag \\
	& & \qquad +  2\Lambda_\mathbf{T}(\sqrt{n_L^c},\mathcal{M}^{q}_{\sqrt{n^c_L},\mathbf{T}}(\hat{M}_{n_L^c})) \Bigg\} 
	+ \frac{A}{\eta} \tau \frac{(\log \mathbf{T})^{10}}{\mathbf{T}} \notag
\end{eqnarray}
Finally, using equations (\ref{appeq:ESJ5}) and (\ref{appeq:EJS1v3}), from \citeauthor[Theorem 6]{lehericy2021nonasymptotic}, we have that for all $\mathbf{T} \geq n_0$, $\tau \geq 1$ and $\eta \leq 1$, with probability at least $1 - e^{-\tau} - 2 \mathbf{T}^{-2}$:
\begin{eqnarray}
	\label{appeq:finalbound}
			D_{KL}^*(\hat{M}_{n_L^c})
	& \leq & \mathbb{NAB}_Q(\tau,\mathbf{T},\sqrt{n_L^c},\mathcal{M}^{q}_{\sqrt{n^c_L},\mathbf{T}}(\hat{M}_{n_L^c})) + f(\mathbf{T})\text{poly}\left((n_L^c)^2\right) \\
	f(\mathbf{T}) & =: & 	\frac{C_{\Lambda}}{\eta} \frac{(\log \mathbf{T})^{17}\log \log \mathbf{T}}{\mathbf{T}}  \notag \\
	\text{poly}\left((n_L^c)^2\right) & =: & ( (m_M - 1) n_L^c + (n_L^c)^2 -  \tilde{m}_M \sqrt{n_L^c}) \notag 
\end{eqnarray}
which corresponds to equation (\ref{eq:EJS6}). \hfill $\Box$

\subsection{Proof of Corollary \ref{cor:volfilter}}

The data generating process appearing in equation (\ref{eq:MarkovSV}) implies that $\Delta y_t$ $\sim$ $\mathcal{N}(0,\bar{V}_t^* )$ conditional on $\mathfrak{U}_t.$ Any assumed estimated model $\hat M$ has $\Delta y_t$ $\sim$ $\mathcal{N}(0,V_t(\hat M) ).$ Let us start with a generic case.
The Kullback-Leibler KL divergence for two univariate Normal distributions, \( P = \mathcal{N}(\mu, \sigma_1^2) \) and \( Q = \mathcal{N}(\mu, \sigma_2^2) \), where both distributions have the same mean \(\mu\) but different variances \(\sigma_1^2\) and \(\sigma_2^2\), is given by:	
	\[
	D_{\text{KL}}(P \parallel Q) = \log\left(\frac{\sigma_2}{\sigma_1}\right) + \frac{\sigma_1^2}{2\sigma_2^2} - \frac{1}{2},
	\]
	since \( \log\left(\frac{\sigma_2}{\sigma_1}\right) \) can also be written as \( -\frac{1}{2} \log\left(\frac{\sigma_2^2}{\sigma_1^2}\right) \) and combining the terms gives:
	\[
	D_{\text{KL}}(P \parallel Q) = \frac{1}{2} \left(\frac{\sigma_1^2}{\sigma_2^2} - 1 - \log\left(\frac{\sigma_1^2}{\sigma_2^2}\right)\right).
	\]
In	addition, since \(\log(1 + x) \approx x\) for small \(x\), in the case where variances are close to each other, the formula is often approximated as:
	\[
	D_{\text{KL}}(P \parallel Q) \approx \frac{1}{2} \left( \frac{\sigma_1^2}{\sigma_2^2} - 1 \right)
	\]
Applying the above to each time period and averaging across the samples yields 
formulas stated in the corollary.
\hfill  $\Box$

\setcounter{section}{0}
\setcounter{equation}{0}
\setcounter{table}{0}
\setcounter{figure}{0}
\setcounter{lemma}{0} 
\setcounter{proposition}{0} 
\setcounter{assumption}{0} 
\setcounter{thm}{0}
\renewcommand{\theequation}{B.\arabic{equation}}
\renewcommand\thetable{B.\arabic{table}}
\renewcommand\thefigure{B.\arabic{figure}}
\renewcommand\thesection{B.\arabic{section}}
\renewcommand{\thesection}{B.\arabic{section}}
\renewcommand{\theproposition}{B.\arabic{proposition}} 
\renewcommand{\theassumption}{B.\arabic{assumption}} 
\renewcommand{\thelemma}{B.\arabic{lemma}} 
\renewcommand{\thethm}{B.\arabic{thm}} 

\section{Some quantum information/computing concepts \label{appsec:QC}}

The basic building block is a Hilbert space $\hilbert$, which is a complete normed vector space over $\complex$ with inner product denoted by 
$\braket \cdot \cdot$ : $\hilbert$ $\times$ $\hilbert \rightarrow \complex$.  In order to describe the states of an $n$-qubit quantum system, it will be of particular interest to consider the tensor product space $\bigotimes_{i=1}^n \hilbert$ and orthonormal basis which will be denoted $\ket{u_i}$ for $i$ = 0, $\ldots$, $n-1$. We are particularly interested in operators $\mathbb{A}:$ $\bigotimes_{i=1}^{n} \hilbert$ $\rightarrow$ $\bigotimes_{i=1}^{n} \hilbert$ that are bounded. An operator $\mathbb{A}$ is bounded if $\|\mathbb{A}\|$ =: $\sup \{ \|\mathbb{A}\psi\| \, | \, \, \ket \psi \in \hilbert \text{ and } \|\psi\| = 1 \}$ $<$ $\infty.$ The set of bounded operators on a Hilbert space $\hilbert$ is denoted by $\mathcal{B}(\hilbert).$ 
Furthermore, it can be shown that $\|\mathbb{A}\|$ equals the absolute value of its largest eigenvalue. In addition to being bounded we also focus on so called Hermitian operators having the property  $\mathbb{A}$ = $\mathbb{A}^\dagger,$ which is the conjugate transpose. We will denote the space of Hermitian operators  by $\mathcal{H}(\hilbert).$ 

\medskip

When a quantum state is precisely determined, it is a pure state, denoted $\ket{v}$ $\in$ $\hilbert,$ with unit norm $\braket v  v$ = 1.\footnote{More precisely, pure states are defined by rays in the Hilbert space space $\hilbert.$ The ray is an equivalence class of vectors that differ by multiplication by a nonzero complex scalar. Vectors within the same ray represent the same pure quantum state. Every ray is represented by a vector . This normalization ensures that the representative vector captures the essential geometric properties and direction of the quantum state.} Given an orthonormal  $\ket{u_i}$ for $i$ = 0, $\ldots$, $n-1,$ a complete description of the quantum system at any given moment in time is via the quantum state: $\ket{v}$ = $\sum_{i=0}^{n-1} a_i \ket{u_i},$ where the coefficients $a_i$ are called the amplitudes and the sum of the squares of their modules is normalized: $\braket v v$ = $\sum_{i=0}^{n-1} {\lvert a_i\rvert }^2 = 1.$ Note that $\ket v \bra{v}$ = $\sum_{i=0}^{n-1} {\lvert a_i\rvert }^2 P_i,$ where $P_i$ $\equiv$ $\ket{u_i} \bra{u_i},$ which is the projection operator associated with $\ket{u_i}.$ The so called Born rule determines that the probability of the system collapsing state $\ket{u_i}$ after measurement equals the square of the magnitude of the amplitude, namely ${\lvert a_i\rvert }^2,$ also equal to ${\lvert\braket{u_i}{v}\rvert }^2$ as well as $\bra{v}P_i\ket{v}$ respectively.

\medskip

A statistical ensemble of pure states, each occurring with a specified probability, is called a mixed state which can be represented as:
\begin{equation}
	\label {mixed state}
\ket \phi =:	\sum_j p_j \ket{v_j}, \quad \sum_j p_j = 1, \quad 0 < p_j < 1 \, \forall j,
\end{equation}
where $p_j$ represent the probability of the system to be in the pure state $\ket{v_j}.$ Measurements of mixed states are mathematically represented by Hermitian operators $\mathbb{A}:$ $\bigotimes_{i=1}^{n} \hilbert$ $\rightarrow$ $\bigotimes_{i=1}^{n} \hilbert.$\footnote{The space of pure and mixed states forms a \textit{convex} subspace within the Hilbert space $\hilbert.$ The extreme points of this subspace are the pure states, whereas mixed states are interior points of the convex subspace. Measurement operators are also sometimes called observables. See \cite{ghysels2024quantum} who present quantum decision-theoretic foundations of measurement operators and mixed states in the context of uncertainty and ambiguity about asset pricing models.}

\medskip

When the system is in the mixed state $\ket \phi$ as in (\ref{mixed state}), then the expected value of an observable, i.e.\ Hermitian measurement operator  $\mathbb{A}$ from the ensemble is defined as:
\begin{equation}
	\label{appeq:expectation}
	\langle \mathbb{A} \rangle_\rho = \sum_j p_j \bra{v_j}M\ket{v_j}   =  \text{tr} \bigl(\mathbb{A}\rho \bigr),
\end{equation}
where $\rho$ represents the density matrix of the system: $\rho$ = $\sum_j p_j \rho_j,$
and $\rho_j$ denotes the density operator of a pure state $\ket{v_j}$:	$\rho_v$ = $\ket{v_j} \bra{v_j}.$
The density operator $\rho$ $\in$ $\mathbb{B}(\bigotimes_{i=1}^{n} \hilbert)$ with the following properties: (a) $\rho$ is Hermitian, 
(b) $\rho$ is a non-negative definite: $\bra{v}\rho\ket{v} \geq 0$, $\forall \ket{v},$ (c) $\rho$ has unit trace: $\text{tr}(\rho) = 1,$ and (d) the eigenvalues $  \lambda_1,\, \cdots,\,\lambda_N  $ of $\rho$ form a probability distribution. 
We will denote the space of density operators by $\mathcal{D}(\hilbert)$.\footnote{Note that
the space $\mathcal{H}(\hilbert)$ is a linear vector space, but $\mathcal{D}(\hilbert)$ is not.
A general linear combination of density operators is Hermitian, but its trace is not necessarily equal to one.} 

\medskip

A quantum operation $\mathcal{O}$ is a linear map:
$\mathcal{O}:$ $\mathcal{B}(\hilbert)$ $\rightarrow$ $\mathcal{B}(\hilbert)$
with the following properties: (a) $\mathcal{O}$ preserves convex combinations of density operators, (b) it is trace preserving: tr($\mathcal{O}\beta$) = tr($\beta$) $\forall$ $\beta$ $\in$ $\mathcal{B}(\hilbert),$ and (c) $\mathcal{O}$ is a completely positive map, i.e.\ $\rho \geq 0 \rightarrow \mathcal{O}\rho \geq 0$, and $\left( \mathcal{O} \otimes I_K \right)$ is positive  $\forall K > 0$.

\subsection{Composite systems \label{appsec:composite}}

The material in this subsection is based on \cite{markov2022implementation} and forms the foundation of the operator-theoretic representation of QHMM.
To represent a quantum system composed of two subsystems, we use a quantum operation that combines the components through a  tensor product state where the subsystems are not entangled: $\rho_L$ $\mapsto$ $\rho_L \otimes \rho_O$ := $\rho_{LO},$ defined on  $\hilbert_{LO}$ = $\hilbert_L \otimes \hilbert_O.$ The partial trace operator applied to a composite state $\rho_L \otimes \rho_O$ with $\rho_O$ = $\ket{u_0} \bra{u_0},$ $\ket{u}$  $\in$ $\{\ket{u_i^O}\}$ yields:
\begin{equation}
	\label{appeqn:partial_trace}
		\mathcal{O}\rho_{LO}  = \text{tr}_O \left( \rho_{LO} \right)  = \text{tr}_O \left( \rho_L \otimes \rho_O \right)  = \sum_{i=0}^{n_O - 1} \left(I_{n_L^q} \otimes \bra{u_i^O} \right)\left( \rho_L \otimes \rho_O \right) \left( I_{n_L^q} \otimes \ket{u_i^O} \right) = \rho_L.
\end{equation}
Let	$\mathcal{O}_U \rho_{LO}$ = $U \rho_{LO} U^{\dagger},$ for unitary operator $U,$ characterizing the hidden Markov chain dynamics, as in Definition \ref{def_unitary_qhmm}. Then the evolution of the hidden state is described by tracing out the $O$ component:
\begin{eqnarray}
	\label{appeqn:op_sum}
	\text{tr}_O \left(U \rho_{LO} U^{\dagger} \right) & = & \sum_{i=0}^{n_O - 1} I_{n_L^q} \otimes \bra{u_i^O} \left[ U \left( (I_{n_L^q} \otimes \ket{u_0}) \rho_S (I_{n_L^q} \otimes \bra{u_0}) \right)
	U^{\dagger} \right] I_{n_L^q} \otimes \ket{u_i^O} \notag \\
& := & \sum_{i=0}^{n_O - 1} K_i \rho_S K_i^{\dagger},
\end{eqnarray}
where the operators $\{K_i\}$, known as Kraus operators are defined as follows:
\begin{equation}
	\label{appeqn:kraus_op}
	K_i = \left( I_{n^q_L} \otimes \bra{u_i^O} \right) U \left( I_{n^q_L} \otimes \ket{e_0} \right).
\end{equation}
The Kraus operators depends not only on the unitary $U$ but also the orthonormal basis  $\{ \ket{u_i^O} \}$ of the emission system. The latter can be rotated with any unitary transformation, and therefore Kraus operators are not unique. 
A mixed state of one of the components in an entangled bipartite pure state can also be defined by performing a projective measurement on the other component. The measurement outcome on one component provides information about the state of the other component due to the entanglement-related correlation between the components. 
Let $\left\{\ket{u_i^O}\right\}$ be an orthonormal basis of the emission component. The corresponding set of orthogonal projectors is: 
$\{ M_i = \ket{u_i^O} \bra{u_i^O}, i = 0, \ldots, n^O -1 \}$ and define the set to operators $\{ P_i \}$ acting on the bipartite system: $P_i$ = $I_{n_L^q} \otimes M_i.$ The measurement outcome in this case has a probability characterized as:
\begin{eqnarray}
	\label{appeq:measurementprob}
	\mathcal{P} \left[ \ket{u_i^O} \mid U \rho_{LO} U^{\dagger} \right] & = & \text{tr} \left( P_i U (\rho_L \otimes \rho_O) U^{\dagger}
	P_i^{\dagger} \right) \notag \\
	& = &   \text{tr} \left( K_i \rho_L K_i^{\dagger} \right).
\end{eqnarray}
Since we are interested in a time series of measurements $\mathfrak{y}_t$ for $t$ = 1, $\ldots,$ $\mathbf{T},$ with associated operators $M_t$ corresponding to the projection onto the observed state at time $t,$ starting from an initial state $\rho^L_0,$ we have the following probabilities:
\begin{align}
\label{appeqn:evolution_measurement_multi}
\mathcal{P}\{\mathfrak{y}_1, \ldots, \mathfrak{y}_{\mathbf{T}}\} = \text{tr}\left\{M_{\mathbf{T}}\left(\sum_i{K_i{\dots M_{2}\left(\sum_i{K_i{M_{1}\left(\sum_i{K_i\rho_0^LK_i^\dagger}\right) M_{1}^\dagger}K_i^\dagger}\right) M_{2}^\dagger\dots}K_i^\dagger}\right) M_{\mathbf{T}}^\dagger\right\}
\end{align}
which is the basis for the likelihood characterization in equation (\ref{eq:log-likeBinyQ}). Similar expressions can be found in \cite{adhikary2020expressiveness} equation (9) and \cite{markov2022implementation} equation (46). Moreover, let $\rho^L_t$ be the mixed state statistical ensemble of the hidden state after having observed a history of observations up to time $t.$ Then, the hidden state statistical ensemble update at time $t+1$ given the $\rho^L_t$ is characterized as:
\begin{equation}
	\label{appeq:quantfilter}
	\rho_{t+1}^L = \frac{\left(\sum_i{K_i{M_{t+1} \rho^L_t M_{t+1}^\dagger}K_i^\dagger}\right)}{\text{tr}\left(\sum_i{K_i{M_{t+1} \rho^L_t M_{t+1}^\dagger}K_i^\dagger}\right)}
\end{equation}

\setcounter{section}{0}
\setcounter{equation}{0}
\setcounter{table}{0}
\setcounter{figure}{0}
\setcounter{lemma}{0} 
\setcounter{proposition}{0} 
\setcounter{assumption}{0} 
\setcounter{thm}{0}
\renewcommand{\theequation}{C.\arabic{equation}}
\renewcommand\thetable{C.\arabic{table}}
\renewcommand\thefigure{C.\arabic{figure}}
\renewcommand\thesection{C.\arabic{section}}
\renewcommand{\thesection}{C.\arabic{section}}
\renewcommand{\theproposition}{C.\arabic{proposition}} 
\renewcommand{\theassumption}{C.\arabic{assumption}} 
\renewcommand{\thelemma}{C.\arabic{lemma}} 
\renewcommand{\thethm}{C.\arabic{thm}} 

\section{On the Transition Kernel for a Parametric Volatility Diffusion Model \label{appsec:hestontrans}}

We present the discrete time and discrete state approximation to the CIR process for $V_\tau,$ appearing in equation (\ref{eq:Heston}) repeated here for convenience:
\begin{equation}
	\label{appeq:Heston}
	dV_\tau = \alpha(\beta - V_\tau) d \tau + \sigma \sqrt{V_\tau} dW_\tau.
\end{equation}
The closed-form expression for the population transition kernel for a $\Delta/k$ increment is as follows:
\begin{equation}
	\label{appeq:transCIR1}
V_{\tau+\Delta/k} = \frac{Y}{2c}, \qquad
\text{ where } c = \frac{2\alpha}{(1 - e^{-\alpha \Delta/k})\sigma^2}
\end{equation}
and $Y$ is a non-central chi-squared distribution with $4\alpha \beta/\sigma^2$ degrees of freedom and non-centrality parameter $2 c V_\tau e^{-\alpha \Delta/k}$. Formally the probability density function is:
\begin{equation}
	\label{appeq:transCIR2}
f(V_{\tau + \Delta/k };V_\tau ,\alpha ,\beta,\sigma)=c\,e^{-u-v} \left (\frac{v}{u}\right)^{q/2} I_{q}(2\sqrt{uv}),
\end{equation}
where $q$ = $(2\alpha \beta/\sigma^2)-1,$ $u$ = $c V_\tau  e^{-\alpha \Delta/k},$ $v$ = $c V_{\tau + \Delta/k},$ and $I_{q}(2\sqrt{uv})$ is a modified Bessel function of the first kind of order $q.$ 
Moreover, the ergodic distribution of the process equals: 
\begin{equation}
	\label{appeq:CIRergo}
f(V;\alpha,\beta,\sigma)=\frac{b^a}{\Gamma(a)}V^{a-1}e^{-b V},
\end{equation}
which is a Gamma distribution with parameters $a$ (shape parameter) and $b$ ($1/b$ scale parameter) where $b$ = 2$\alpha/\sigma^2$  and  $a$ = $2\alpha \beta/\sigma^2.$

\smallskip

Armed with the non-central $\chi^2$ distribution appearing in equation (\ref{appeq:transCIR2}) we can construct a discrete state Markov chain as follows: (a) we adopt a binning scheme similar to that for the observable returns and call it $\mathcal{B}_V$ := $\{b_1^V, \ldots, b_{n_L}^V\}$  (b)  we associate a spot volatility state $V_i$ with each bin to populate the set $\mathcal{H}_L$ of hidden states, each representing a midpoint of the binned partition, and finally construct the transition density matrix with transition probabilities from state $V_{t}$ = $V_i$ to $V_{t+1}$ = $V_j.$ 

\smallskip

We assign each spot volatility a numerical value using $V_i$ = $F_e\left((i+1)(n^c_L+1)\right),$ 
for all $i \ \in 0, \dots n^c_L-1$ where $F_e$ is the CDF of the ergodic distribution appearing in equation (\ref{appeq:CIRergo}). With an exact value attributed to each spot volatility, we determine the transition probability from $V_\tau$ to $V_i$  
as the difference in the non-central $\chi^2$ transition kernel distribution at the two midpoints adjacent to $V_i$. More formally, the transition probability from $V_i$ to $V_j$ is
\begin{equation}
	\label{appeq:cdf_diff}
	q_j^i = F_{tr}(2c V_{j+}) - F_{tr}(2c V_{j-}) 
\end{equation}
where $V_{j+}$ is the midpoint between $V_j$ and $V_{j+1}$, or infinity in the case of $V_{j\ max}$. Likewise,  $V_{j-}$ is the midpoint between $V_j$ and $V_{j-1}$, or negative infinity in the case of $V_{0}$. We use equation (\ref{eq:cdf_diff}) to populate the tightly parameterized transition matrix
\begin{equation}
	\mathscr{A} = \left[
	\begin{array}{cccc}
		q^0_0 & q^0_1 & \dots & q^0_{n^c_L-1} \\
		q^1_0 & q^1_1 & \dots & q^1_{n^c_L-1} \\
		\vdots & \vdots & \vdots & \vdots \\
		q^{n^c_L-1}_0 & q^{n^c_L-1}_1 & \dots & 
	\end{array}
	\right].
\end{equation}

\setcounter{section}{0}
\setcounter{equation}{0}
\setcounter{table}{0}
\setcounter{figure}{0}
\setcounter{lemma}{0} 
\setcounter{proposition}{0} 
\setcounter{assumption}{0} 
\setcounter{thm}{0}
\renewcommand{\theequation}{D.\arabic{equation}}
\renewcommand\thetable{D.\arabic{table}}
\renewcommand\thefigure{D.\arabic{figure}}
\renewcommand\thesection{D.\arabic{section}}
\renewcommand{\thesection}{D.\arabic{section}}
\renewcommand{\theproposition}{D.\arabic{proposition}} 
\renewcommand{\theassumption}{D.\arabic{assumption}} 
\renewcommand{\thelemma}{D.\arabic{lemma}} 
\renewcommand{\thethm}{D.\arabic{thm}} 

\section{Details Empirical Implementation \label{appsec:implementation}}

As specified in Section \ref{sec:qinspired} we study two classes of classical models: the parametric and non-parametric cases. The two types of models differ in terms of the parameters $\theta$ that they use to construct a spot volatility transition matrix. Here we outlines the procedure for each case. The simpler of the two methods in the non-parametric classical model, where we use the $n^c_l \left(n^c_l - 1 \right)$ parameters $\theta$ to construct a spot volatility transition matrix $\mathscr{A}$ which takes the form
\begin{equation}
	\mathscr{A} = \left[
	\begin{array}{cccc}
		\theta_0 & \dots & \theta_{n^c_l-1} & 1 - \sum_{i=0}^{n^c_l-1} \theta_i \\
		\theta_{n^c_l} & \dots & \theta_{2 \left(n^c_l-1\right)} & 1 - \sum_{i=n^c_l}^{i=2 \left(n^c_l-1\right)} \theta_i \\
		\vdots & \vdots & \vdots & \vdots \\
		\theta_{\left(n^c_l-1\right)^2}	& \dots & \theta_{n^c_l  \left(n^c_l-1\right)} & 1 - \sum_{i=\left(n^c_l-1\right)^2}^{i=n^c_l \left(n^c_l-1\right)} \theta_i
	\end{array}
	\right],
	\label{appeq:transitionmatrix}
\end{equation}
and assume that the initial latent state is drawn from the ergodic density of $\mathscr{A}$. We solve for this initial state analytically, however for $n_c^l$ this could start in any state with a pre-sample burn-in period. 

\medskip

For the parametric case, we present the discrete time and discrete state approximation to the CIR process detailed in Appendix Section \ref{appsec:hestontrans}. 

Armed with the non-central $\chi^2$ distribution appearing in equation (\ref{appeq:transCIR2}) we can construct a discrete state Markov chain as follows: (a) we adopt a binning scheme similar to that for the observable returns and call it $\mathcal{B}_V$ := $\{b_1^V, \ldots, b_{n_L}^V\}$  (b)  we associate a spot volatility state $V_i$ with each bin to populate the set $\mathcal{H}_L$ of hidden states, each representing a midpoint of the binned partition, and finally construct the transition density matrix with transition probabilities from state $V_{t}$ = $V_i$ to $V_{t+1}$ = $V_j.$  We assign each spot volatility a numerical value using
\begin{equation}
	V_i = F_f\left(\frac{i+1}{n^c_l+1}\right) 
\end{equation}
for all $i \ \in 0, \dots n^c_l-1$ where $F_f$ is the PDF of the ergodic distribution in equation (\ref{appeq:CIRergo}). With an exact value attributed to each spot volatility, we determine the transition probability from $V_\tau$ to $V_i$  
as the difference in $F_Y$, the CDF of the non-central $\chi^2$ distribution in equation (\ref{appeq:transCIR2}) at the two midpoints adjacent to $V_i$. More formally, the transition probability from $V_i$ to $V_j$ is
\begin{equation}
	\label{eq:cdf_diff}
	q_j^i = F_Y(2c V_{j+}) - F_Y(2c V_{j-}) 
\end{equation}
where $V_{j+}$ is the midpoint between $V_j$ and $V_{j+1}$, or infinity in the case of $V_{j\ max}$. Likewise,  $V_{j-}$ is the midpoint between $V_j$ and $V_{j-1}$, or negative infinity in the case of $V_{0}$. We use equation (\ref{appeq:cdf_diff}) to fill out the tightly parameterized transition matrix
\begin{equation}
	\mathscr{A} = \left[
	\begin{array}{cccc}
		q^0_0 & q^0_1 & \dots & q^0_{n^c_l-1} \\
		q^1_0 & q^1_1 & \dots & q^1_{n^c_l-1} \\
		\vdots & \vdots & \vdots & \vdots \\
		q^{n^c_l-1}_0 & q^{n^c_l-1}_1 & \dots & q^{n^c_l-1}_{n^c_l-1} \\
	\end{array}
	\right].
\end{equation}
Each procedure results in a spot volatility transition matrix, and associated volatility values. The parameters used to construct said matrix are the distinguishing feature between the parametric and non-parametric cases.

\medskip

With our spot volatility transition matrix and bin value, next we calculate the emission matrix using the same procedure for both classes of model. The critically, the complexity of this process scales with $\mathcal{O}\left(k^2\right)$, where $k$ is the number of spot volatilities per integrated volatility time step. With high frequency spot volatilities we see this process as the primary computational bottleneck. We deconstruct the process of deriving the emission matrix into three distinct computational tasks. First, we solve $g(\bar{V}_j | V_t; M_{n^c_l})$ by calculating the probability of all $k$ step spot volatility sequences $\bar{S}_j$. We add the probability of $S_j$ to an approximate function $\tilde{g}(\bar{V}_j | V_t; M_{n^c_l})$ for each $V_i \in S_j$ and the $\bar{V}_j$. We determine the index of the integrated volatility using
\begin{equation}
	j = -k + \sum \ i \ \forall \ V_i \ \in S_j.
\end{equation}
After the data from every sequence has been added to $\tilde{g}(\bar{V}_j | V_t; M_{n^c_l})$, we normalize the result to yield the probability
\begin{equation}
	g(\bar{V}_j | V_t; M_{n^c_l}) = \frac{\tilde{g}(\bar{V}_j | V_t; M_{n^c_l})}{\sum_{j'=0}^{j_{max}} \tilde{g}(\bar{V}_{j'} | V_t; M_{n^c_l})}.
\end{equation}
In other words, for a given spot volatility, the probabilities of each integrated volatility sum to one. In our example analysis we use $k=4$ spot volatility time steps for each integrated volatility. The time complexity of this dynamic programming problem scales with $\mathcal{O} \left(n^c_l k^2\right)$. 

Next, we calculate the probability of an emitted state based on the integrated volatility. For this we separate the rate of return over the chosen time period into $n_O$ distinct observation bins. We scale the time series data such that the values of each integrated volatility bin is simply the average of the spot volatilities over the course of the sequence. The probability of its each observation during a sequence with integrated volatility $\bar{V}_i$ is determined by the cdf of $\mathcal{N}\left(0, \bar{V}_t\right)$. The time complexity of calculating $P\left(x|\bar{V}_{t+1}\right)=\bar{V}_j$ in this way scales with $\mathcal{O} \left(n^c_l \ k \ n_O \right)$.

Finally, we determine the emission probabilities for a given spot volatility by performing the summation in equation \ref{eq:dirac} over all values of $\bar{V}_j$. The python implementation of this procedure is available in the linked github repository \cite{Morgan_QHMM_MLE_2025}. The result is a matrix that approximates emission probabilities given lower frequency observations which are dependent on a higher frequency latent Markovian process. We use the same procedure for both the tightly and loosely parameterized diffusions.

\end{document}